# What's beyond Presence? Dimensionality, Control and Information Spaces

Eugene Ch'ng
School of Cultural and Creativity
BNBU Centre for Computational Culture and Heritage | NVIDIA DLI
Beijing Normal University-Hong Kong Baptist University
eugenechng@bnbu.edu.cn

**Abstract**

What's after presence? Spatial presence, the sense of 'being there,' will become less of a primary objective; instead, it will become a baseline expectation of VR. More than six decades after its invention, virtual reality is evolving from a technical endeavour into a cultural, social, and phenomenological medium, offering experiences that can be considered distinct modes of reality. Existing theories focused on perceptual illusions are insufficient to evaluate the depth of these emerging experiences. A framework that guides the design and assessment of immersive environments, identifying key technical and abstract dimensions afforded by the virtual environment, has become necessary. These dimensions include spatial, placeness, temporal, social, cultural, cognitive, and psychological parameters. The central argument of the article is that virtual environments should move beyond the technical dimension to explore other information channels that can enhance the user's experience. This shift in focus from presence to the orchestration of experience invites creators beyond the technical fields into the design, development, and evaluation of meaningful immersive worlds.

**Keywords:** presence, immersion, VR, theory, framework, dimensionality, control, information spaces

## 1. Introduction

This article stands on the thesis that the sense of 'being there' will increasingly become a baseline expectation rather than the primary goal of VR. As technical immersion establishes the foundation to information channels, the focus should shift from creating environments that feel real to shaping the user's experience and directing their journey within it. In short, the argument is that we should move beyond just asking "Does it feel real?" towards "What specific feeling, interpretation, or outcome are we trying to evoke, and how can the medium be modulated to achieve it?" These information channels such as spatial, temporal, placeness, social, cultural, cognitive, and psychological can be conceptualized as non-technical, abstract dimensions for achieving designer-defined phenomenological goals. The designer's aim is to shape the user's lived experience, encompassing how they perceive, feel, interpret, interact, and make meaning within the virtual environment.

The core scholarly debate on presence has centred on its definition as a psychological state, alongside investigations into spatial, social, and narrative presence. For decades, research has leaned heavily towards investigating the factors that contribute to 'immersion' and its relation to 'presence'. These range from performance of technological and media properties, such as tracking and display fidelity, that influence user perception. This lengthy inquiry is driven by the need to understand the presence itself and its significance in specific uses, such as domain-specific task performance, learning, and greater emotional engagement.

Beyond presence research, the community has made significant progress in creating virtual environments with intended outcomes; however, numerous examples of VR systems still fall short of their intended goals. We have encountered virtual environments that offer limited fidelity or functionality, often resulting in poor replications of the physical environments they aim to simulate. Simple configurations of sensorimotor components, such as appropriate interaction and navigation, were not implemented properly. This basic struggle with technological immersion is detrimental towards the affordance of additional properties of the medium, such as the inclusion of dimensions that can significantly affect cognitive and psychological processes. These limitations highlight the ongoing challenges of harnessing the unique dimensionality that immersive technologies afford. Many VR applications have robust sensorimotor engagement but lack narrative design. Others have not taken advantage of the freedom offered by virtual reality in the 3D Cartesian space. Instances such as virtual classrooms that unnecessarily replicate physical settings, even though VR is not constrained by the physical need to sit on chairs or face rectangular displays. There appears to be a growing number of VR applications that are underutilized or fail to deliver meaningful experiences. I suggest that this may stem from a lack of a guiding framework.

The question of the need for presence depends on the uses of the virtual environment. There are suggestions that task performance can serve as an objective measure of presence (Kalawsky et al., 1999; Schloerb, 1995), and that a positive correlation exists between the two (Witmer & Singer, 1998). However, if an environment is task-performance as a function of spatial movement and navigation within simulations of the physical environment, then presence is likely to affect performance. The inverse relationship between presence and performance is demonstrated in instances where presence decreases in relation to performance increases, particularly when information clutter is removed from the environment, as seen in air traffic control displays (Ellis, 1996). In the majority of cases, the goal of creating a virtual environment that replicates a real physical place or a fictional scene is to enable users to experience what the designers intended for them, in which case, the sense of 'being there' is the baseline expectation.

Despite extensive research on presence, current frameworks do not adequately explain how designers can orchestrate experiences that integrate cultural, social, narrative, cognitive, and psychological factors. This article proposes a comprehensive, multidimensional model for designing VR experiences that leverages dimensionality, control, and information flow. In brief, the article contributes to the field by:

- reframing presence as foundational rather than aspirational in light of technological advances;
- proposing a comprehensive, multidimensional framework for VR experience design;
- integrating cultural, narrative, cognitive and psychological dimensions into existing VR theory;
- introducing actionable design 'dials' as control that modulate effects, and translate theory to practice;
- positioning VR as an information-rich medium that requires the orchestration of information flow.

## 1.1 The beginnings of presence

Virtual Reality has a long history of development, having gone through cycles of hype and disillusionment, only to resurge with new technological advancements, culminating in a cycle of despair. Is virtual and augmented reality leading anywhere useful? From its early inception in the 1960s with elaborate devices such as Morton Heilig's Sensorama, and Ivan Sutherland's stereoscopic Sword of Damocles, the field has repeatedly captured public imagination, yet, apart from enterprise use such as military simulations and flight simulators, VR has struggled with limitations in hardware, affordability, and usable content that can lead to mass adoption. Why is VR evading mainstream use? Have we not properly understood immersive media and how to utilise it to its full potential? Is technology at fault, or is it that contemporary society is not yet ready for virtuality? Surely the source of the problem lies in the design of the virtual environment, rather than the users?

The 1990s brought optimism with entertainment devices such as Nintendo's Virtual Boy, but the technology fell short of providing a truly immersive experience. The launch of *Presence: Teleoperators & Virtual Environments*, now *Presence: Virtual and Augmented Reality* occurred in that period. As evident in Figure 1, many of the topical research articles in current virtual reality (VR) and augmented reality (AR) originated from the journal's first year of publication, in 1992.

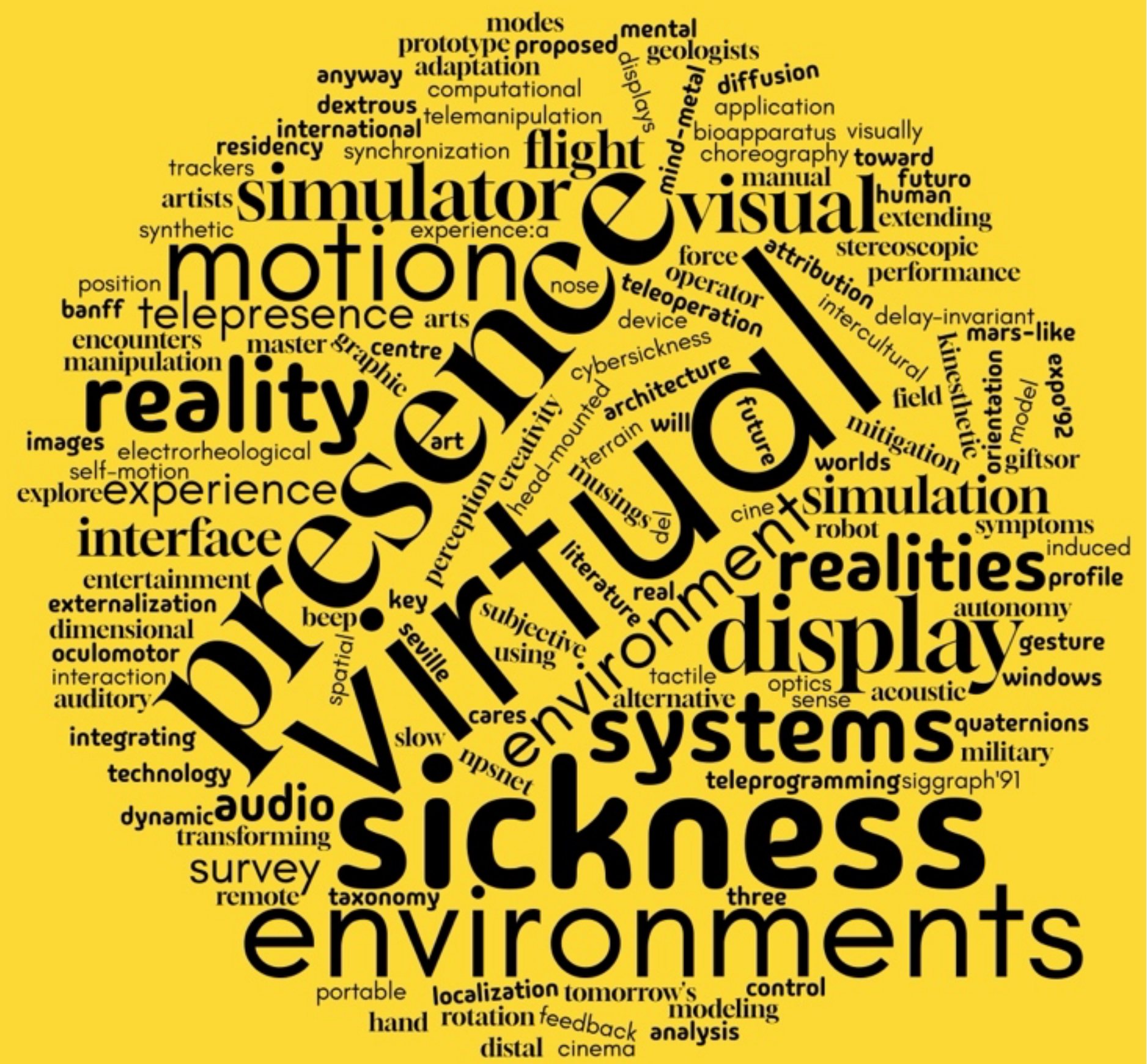

Figure 1. The Presence journal's complete 1992 volume and six issues of article titles generated as a word cloud, highlighting larger keywords that appear more frequently within the pool of keywords that cover a broad range of topics.

Even though the computer graphics capability of the 1980s VR were primitive, early researchers and developers investigating the technology saw broad ranging utilities, as Biocca and Levy have noted in 1991, ''Introductory VR books often describe virtual reality as the next logical step in the history of communication media, and a Delphi panel survey predicts that communication applications of virtual reality will amount to more than 60% of the marketplace when the technology matures' 2013). The predicted 60% did not occur, and 2013 came and went; VR was not at all in the marketplace. This, however, demonstrated the community's imagination at the time and the technology's untapped potential. The founding fathers of Presence and the journal's broad beginnings may be a strong reason why the field as a whole has sustained this long.

The late founding editor-in-chief Nat Durlach studied a 'broad range of factors, from haptics, audition, vision, somatosensation, locomotion, wayfinding, display technology, robotics, telerobotics, and user interfaces, to motion sickness.' (Lackner, 2016). The Presence Forum Article *Nat Durlach and the Founding of Presence* (p.161) highlighted Nat's thinking, 'Nat realized that virtual reality was not only a tool for training and familiarization and rehabilitation but also a way to gain greater knowledge and insights into the nature of our consciousness as sentient beings situated in an environment. He realized that the notion of 'presence', of feeling situated in a particular environment or context, was something that our central nervous system (CNS) constructs based on sensory-motor and cognitive information received about our current state.' In *Remembering Nat Durlach (Slater, 2016)*, Mel Slater, co-editor-in-chief of Presence between 2011 to 2014, known for his contribution to VR, and earlier works when Nat was editor-in-chief (Slater, 1999; Slater et al., 1994a; Slater & Wilbur, 1997a) gave testament to Nat as 'one that concentrated more on the human factors side than the systems and algorithms side.' This was how Presence began.

### 1.2 Technological momentum

At the turn of the century, the four crucial technologies and the four auxiliary technologies proposed in Brook's 1999 article *What's Real About Virtual Reality* (Brooks Jr, 1999), which reviewed the fields progression, were all based on technology and systems development, and yet, these proposals were all purposed for the VR experience of preexisting human experience in physical scenarios. VR required more than just technology and system development, but the technical foundations of sensorimotor fidelity were inadequate.

After a period of silence in the early 2000s, and following the launch of the Oculus Rift, HTC Vive, and PlayStation VR in the mid-2010s, VR technology rapidly gained traction, attracting substantial investment from major tech giants, who invested resources and fueling technological advancements in the field. These efforts collectively laid the groundwork for an increasingly interconnected digital landscape, paving the way for the emergence of an envisioned metaverse. This highlights a shift toward technologies targeting full technical immersion (see, for example, Choi et al., 2025; Geng et al., 2025; Wong et al., 2017; Xiao et al., 2025). The push towards this new vision was more ambitious than the VR community had envisioned.

As these technologies converged, the concept of the metaverse evolved from science fiction (Stephenson, 1994) into a nascent tangible digital frontier, envisioning a large interconnected digital ecosystem that was expectant of a future of work and leisure requiring not only our attention, but 'our presence' rather than our 'sense of presence'.

### 1.3 Three decades of presence: technology vs. human experience

In view of recent developments and based on an attempt to classify Presence articles from 1992 and 2025, a distinct but converging polarity emerges. Most articles can be categorized either under Technology and Systems Design (TSD) or Human Experience and Perception (HEP). This is a natural category, as VR encompasses both technology and systems, as well as the digital aspects of the immersive medium that bring users perceptually into another space, thereby involving the human experience and perception. As Dooley and Skarbez put it, "Presence is perhaps the defining feature of VR experiences: This journal did not acquire its name by accident." (Murphy & Skarbez, 2020).

Figure 2 is a scatter plot of the appearances of keywords of all Presence article titles (1992-2025) using the ScatterText library. Article titles were preprocessed by removing stop words, non-ASCII characters, empty strings, and unwanted phrases related to editorials, reviews, etc. The remaining keywords were tokenized and plotted as frequency of occurrences in the x (Human Experience and Perception) and y (Technology and Systems Design) axes. Keywords appearing more frequently in the TSD class would be pushed to the top in the y-axis, and those appearing in the HEP class titles would be pushed to the far right in the x-axis.

Consequently, the convergence at the peak frequencies of the two categories would be keywords familiar to our community, and these are 'virtual', 'reality', 'environments', 'immersive', 'real', 'visual', 'haptic', and 'control'.

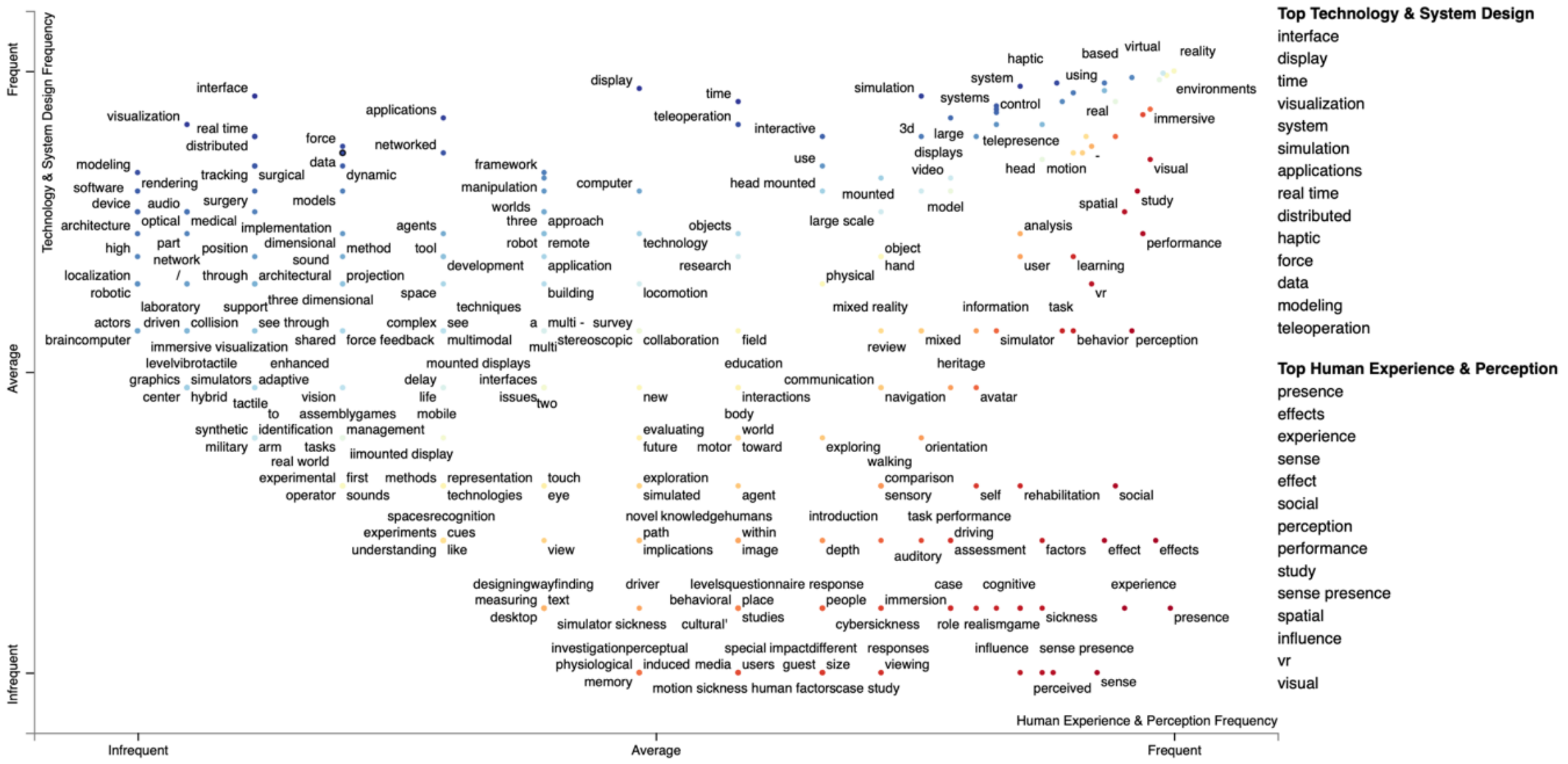

Figure 2. Scatterplot of keyword frequencies appearing within the titles of all Presence articles (1992-2025) apart from editorials, reviews, remembrances, special issues and unclassed article titles. Technology and System Design (document count: 499; word count: 3,640), Human Experience and Perception (document count: 587; word count: 4,828).

Presence journal leans toward the human experience and perception at 54% of all published articles. I projected that this trend would continue, as more case studies are published within the journal, and beyond, in other disciplines. This is indicative of the broader field, and the journal, of course, does not capture all topical articles which are distributed across other similar journals. However, I believe that all technological and systems development for VR, AR and MR will lead to the support of the human experience within virtual and augmented spaces. If we consider the use cases for VR, we would see that they would fall within the centrality of the end-user experience if mass adoption were to occur. These use cases would include entertainment, communication, education, health and well-being, design, and the sociality of remote collaboration and togetherness within virtual spaces. These are areas where the envisioned metaverse is expected to focus. These would be an illusory space (Biocca & Levy, 2013), 'a mutually accepted make-believe space, a "consensual hallucination," where the fiction, game, or entertainment takes place "long ago and far away".

VR and AR researchers are Gibsonian (Gibson, 1966). The projection of the VR field does not veer from its original goals; rather, as the field advances, there is a noticeable shift toward research centred on human experience and perception within virtual and augmented environments. I believe that as the technology matures, the human experience and meaning-making become increasingly central to research in these spaces. This evolution invites the incorporation of theoretical frameworks from the arts, humanities and social sciences, providing critical lenses through which virtual environments can be interpreted and understood. Disciplines such as philosophy, cultural studies, media and communication, and critical theory would gain prominence. Aesthetics would guide the design of immersive storytelling and experiences, while ethical and philosophical inquiries into identity, autonomy, and reality would become increasingly critical. While hardware and software develop in parallel, the ultimate convergence of the field will be centered on the human experience within an envisioned metaverse, where all contents and communication are conceptualised as information flows (See section 3.3.9), and where effective and meaningful information is transmitted. Indeed, it is information that was described with regards to representations, social interaction and interpersonal communication (Bailenson et al., 2008), virtual environments are referred to early as 'an organisation of sensory information (Blascovich, 2002)'.

An envisioned metaverse will emerge as both singular and a connected web of virtual worlds, with an underlying framework for communication and exchange of value, and as an extension of our already-mediated existence, driven by big tech's economic incentives, democratized 3D and content generation tools, and the provision of XR technology's dimensionality and possibilities. Together with Generative AI, virtual spaces and contextual content would be dynamically created and populated on demand. While challenges remain, these converging trends make their emergence inevitable.

Technological advancements have accelerated in the past few decades, but the conceptual frameworks guiding the field have remained largely unchanged. Perhaps it is time to explore new frameworks from which we can begin to understand the emerging landscape.

### 1.4 Information afforded by dimensionality

Immersive media offers modes of communication that transcend the two-dimensional plane of traditional formats. By leveraging the 3D spatial dimensions, these technologies create new affordances for content design that make use of other dimensions. This additional dimensionality enables information to be presented in more intuitive and meaningful ways, influencing users in ways that were previously impossible in 2D media. For example, complex data, architectural designs, or training simulations can be experienced and understood more effectively when users can interact with them in a three-dimensional space. Reading history from a book would not provide the experience of being in the recorded place. Similarly, visiting a museum and viewing relics in enclosed glass panels would not be the same as interacting with those objects in the context of where they were a thousand years ago. It is extremely hard for even those with highly developed imaginations that would allow them to experience an unfamiliar cultural setting through cognitive processes. It would be extremely difficult to read a textbook on the life of a community in a South East Asian fishing village and understand what they have experienced, for it is the mind that interprets and imagines the environment. It would be inadequate to watch a documentary on those displaced by war and claim understanding, for what can be portrayed is a framed view by the director, the best effects one can achieve with the medium from which the documentary was created. We are not present, nor there in the environment. A 360° documentary, even though it is not actively navigable, presents an additional dimension (weak representation is engaged, see section 2.3), one in which users can feel 'present' on a boat building site, on a boat swayed by the current, and have the full view of the environment when nets filled with different species of marine life are hauled onboard. Even without the sensation of temperature, humidity, olfaction, immersion, and the spatial sound of the 360° experience, the cross-stimulation of other sensations, such as bodily chills through imagination, would still occur. Meaning that the richness of information is afforded by the extra-dimensionality. Extra dimensionality, Cartesian in this sense, can give rise to the generation of additional dimensionality (information channels) that engages with strong representations – a dimension gives rise to other dimensions. In Cai et al.'s research (Cai et al., 2018), for example, a reconstructed grandma's kitchen from an old village elicited cross-stimulation of olfactory sensation, a product of the imagination, even though the VR system has no olfactory devices. Many such examples exist in virtual reality experiences. At the cognitive and psychological level, this aspect of how user imagination can complement the information provided by the system is rarely discussed, following the presentation of the three I's of VR – immersion, interaction, and imagination (Burdea & Coiffet, 2003). In fact, there has been little mention of imagination, referring to 'the capacity of our brain to fill gaps in the imperfect sensorial information it receives.' (p.3, Chapter 1), after the article, and in contrast with the many articles written on interaction and immersion. A decade after the influential work of Burdea and Coiffet (2003), a nuanced definition was proposed that highlights the role of cues from a medium in shaping imagination. According to Biocca and Levy, VR is characterized by the 'The replacement of everyday sensory reality for user-generated illusions driven by cues from a medium: for example, words of the storyteller; a storybook picture; and an action-packed, car chase; and so forth.' (Biocca & Levy, 2013). The thesis is that the construction of VR experiences relies on the user's imagination and agency.

Within the reality-virtuality continuum (Milgram & Kishino, 1994), extending the environment toward true virtual reality, where users have full freedom of interaction and navigation in a fully immersive environment, increases the amount of information the space can afford. This, in turn, expands the range of combinations and permutations through which the medium can communicate information. Conversely, moving a step closer to the real-world end of the continuum by enhancing physical reality with contextual information via spatial computing, practical AR headsets, and augmented digital content, brings users closer to utility and real-world applicability, where abstract dimensions can build on.

There is a growing need for a new conceptual framework that moves beyond the foundational but now limited paradigms of early VR theories. The "Three I's" of VR – immersion, interaction, and imagination offered an initial understanding of what virtual reality affords, while the reality-virtuality continuum mapped the spectrum of systems positioned between the real and the virtual. However, these concepts no longer provide sufficient insights into the evolving dynamics of current information-rich immersive environments. They serve primarily as descriptors rather than drivers for enquiry. Concepts developed since the 1990s – such as spatial presence, social presence (Biocca et al., 2003; Mennecke et al., 2010; Rettie, 2003), flow (Csikszentmihalyi & Csikzentmihaly, 1990), plausibility illusion (Skarbez, Neyret, et al., 2017a, 2017b; Slater, 2009), visual fidelity, authenticity and realism (Gilbert, 2016), along with theories drawn from adjacent disciplines, have provided valuable frameworks for assessing the effects of user perception and experience. Building upon these foundational theories, they continue to inform our understanding of quality and engagement in immersive environments. As extended reality technologies become increasingly complex and entangled with everyday life, perhaps there is value in expanding these perspectives into a framework from which we may make further progress.

# 2. What's After Presence?

## 2.1 Spatial presence via sensorimotor immersion

Historically, the dominant argument has been that increases in sensorimotor immersion directly transport users into a state of presence (Barfield et al., 1995; Barfield & Weghorst, 1993; Biocca, 1999; Draper et al., 1998; Held, 1992; Lombard, 2000; Lombard et al., 2000; Lombard & Ditton, 1997; Sheridan, 1992; Slater et al., 1994b; Steuer, 1992).

The relationship between 'immersion' and 'presence', therefore, is how well the system can deliver "displays (in all sensory modalities) and tracking that preserves fidelity in relation to their equivalent real-world sensory modalities. The more that a system delivers displays (in all sensory modalities) and tracking that preserves fidelity in relation to their equivalent real-world sensory modalities, the more that it is 'immersive'." (Slater, 2003). In this simple model, immersion is what technology provides.

But is presence solely a function of sensorimotor immersion? The answer is more complex. Presence in a mediated environment is a product of both objective technological factors and the user's subjective internal state. It is determined by both media characteristics (external, objective) and user characteristics (internal, subjective) (W. A. IJsselsteijn et al., 2000; Slater & Wilbur, 1997b). This suggests that pure technological definition is incomplete.

Presence is better understood as a human reaction to immersion. In other words, immersion is what technology does, and presence is what we feel. Presence is therefore also subject to individual reactions toward immersion (Camci, 2019; Green & Brock, 2003; Pfister & Ghellal, 2018; Sacau et al., 2008; Slater & Usoh, 1993; Spennemann & Orthia, 2022), including underlying mental health conditions (Huang & Alessi, 1999). The sense of presence therefore, cannot be a simple binary state. Furthermore, Heeter (Heeter, 2003) has also stated "Presence is a series of moments when cognitive and perceptual reactions are closely tied to current sensory impingements". The reported level of presence varies considerably over time, depending on the extent and naturalness of the sensory information available in the stimulus material (W. IJsselsteijn et al., 1998; W. A. IJsselsteijn et al., 1997).

The subjective experience has also been the subject of a lengthy debate. Early attempts described it using the phrase 'suspension of disbelief' (Higgins, 1998; Murray, 2012; Walton, 1980) to refer to the notion that presence requires belief or its suspension, a conscious cognitive act. Others referred presence as an 'illusion of nonmediation' where users failed to acknowledge the role of technology ((ISPR) International Society for Presence Research, 2000; Lombard & Ditton, 1997), and the illusion or feeling of 'being there', as a subjective sense of self-location in a mediated space (Clark, 1998a; Heeter, 1992; Minsky, 1980; Riva et al., 2003; Slater, 2018). The two prevailing meanings of presence thus far are "being there" and "nonmediation" (Skarbez, Brooks Frederick P, et al., 2017).

Challenging this conceptualisation, Dooley and Skarbez (Murphy & Skarbez, 2020) clarified conceptual confusion around presence, stating that presence is not a monolithic construct reducible to suspension of disbelief, illusion of nonmediation (fails to consciously acknowledge that technology was mediating a part of their experience), or merely the experience of 'being there' (being present in a simulated virtual environment). They proposed that the concept is a multi-layered perceptual cognitive phenomenon, with aspects of spatial presence being felt automatically through sensorimotor input and perceptual attention. Place illusion, therefore, is not a conscious influence. It requires no cognitive attention and is separate from rational belief. Spatial presence, therefore, is not willed into existence but is felt automatically.

## 2.2 The 'empty room' thought experiment

Just as we can feel present in a physical empty room, a user can feel a sense of 'being there' in a simulated empty room if the technology convinces the brain that the person is occupying a different space from the physical one. In this case, presence is not about how interesting or exciting the room is; it is about how the person perceives their location in that instance, taking into account weighted factors such as head tracking, parallax, scale, depth, sound, and acoustics. The state of presence need not be linked to an exciting room that is full of things to look at. Standing in a real, empty room might be boring, but we do not doubt that we are there. A well-simulated environment tricks the same sensorimotor cues our brain uses to create that feeling of being there.

Slater (Slater, 2003) referred to an example of a quadraphonic sound system, which plays music, and of someone experiencing as if they are present at the theatre where the orchestra is playing. It does not matter what is being played, nor how interesting, emotionally captivating, or beautiful the music is; it is only the content. The form is more important, and that presence is about form, not the content itself.

Once form is perfected, content becomes the key differentiator. If we were to move beyond presence, then content becomes the key differentiator when high-fidelity sensorimotor immersion is achieved. For what is an empty virtual room even if it feels completely real? As the technology of "form" becomes more standardized and effective, the creativity and meaningfulness of the "content" will define the quality of a VR experience.

Form and content are separated by Slater's argument (Slater, 2003), but they are often intertwined. An interactive object (content) may draw a user's attention and require them to engage their sensorimotor skills, which can, in turn, heighten their sense of presence (a reaction to the 'form'). Alternatively, an emotionally captivating narrative (content) can draw a user's attention away from minor technical flaws in the simulation (form), triggering a strong representation, which leads to a higher sense of presence (narrative).

The significance of a designed experience depends on what the environment enables. Once the sense of 'being there' is achieved, the next question for the user becomes, 'Why am I here, and what can I do?' Spatial presence (physical or virtual) must lead to some intention and action. In being in an empty room that is without distractions, the very first perceptual reaction would be that I am conscious, and that I am aware that I am in a room. The immediate reaction that follows is certainly the introspection of the desire. This invariably leads to the intention to fulfil that desire, and an arbitrary action follows. I may not immediately take any action, but this lack of action is an intention in itself, which leads to the default state of inaction. A room that is filled with interesting objects would lead to the inspection of those objects, if the virtual environment allows such an action. A room without any objects will lack stimulation, causing the user's mental state to drift.

Every virtual environment designed implies a purpose, and the purpose must go beyond the sense of 'being there' to a purposeful experience.

### 2.3 Cognitive and attentional presence

Biocca's 'Book Problem' (Biocca, 2003) challenges the traditional two-pole psychological model of presence, which suggests that people oscillate between physical and virtual spaces, and that increases in sensorimotor immersion are the principal variables influencing the movement from the former to the latter.

Biocca argues that 'a high level of sensorimotor is not a necessary or sufficient condition for presence.' (p.8). People reported high levels of presence while reading books and in the dream state, despite the absence of sensorimotor input (p. 9). Furthermore, despite the maximum level of sensorimotor immersion, people are sometimes not present in their physical environment. Biocca posits that 'books achieve their levels of presence by making heavy use of the imagery space to "fill in" the spatial model cued by the book' (p.9), and that the traditional assumption fails to incorporate the role of spatial attention and mental imagery.

Books achieve their levels of presence by making heavy use of the imagery space to "fill in" the spatial model cued by the book. Biocca adds a third pole, which incorporates self-generated and mental imagery, and proposes that presence results from gravitation among the three spaces: real, virtual, and imagined spaces.

Schubert and Crusius (Schubert & Crusius, 2002) formulated five theses, claiming that the sense of presence is a cognitive construct created from immersive stimuli, and that it is the user's mental models, not just the technology's immersive fidelity that determines the experience of presence, citing Slater et al. (Slater et al., 1994b), "perceptions generated by the [immersive virtual environment] are mediated through the mental models and representation systems that structure participants' subjective experience." Therefore, 'the structure of this mental model determines whether the user experiences a sense of presence or not.' The main claims are that presence is the same across media forms, cognitive mediation is the key and that media differ in spatial involvement and bodily interaction. Schubert and Crusius also believe that books evoke presence through narration, rather than sensory immersion; however, there was no direct explanation provided as to how narratives evoke presence. Schubert and Crusius shifted the weight towards the individual's perception and cognition.

Turner and Turner (Turner & Turner, 2011) contested Biocca's argument that imagery fills in spatial gaps, arguing that presence in books is more than visual or spatial imagery, and inspected cognitive representation (Clark, 1997, 1998b; Clark & Grush, 1999), narrative transport (Green, 2005; Nell, 1988; Ryan, 2015a) and neuroscience of imagination and dreaming (Decety, 1996; Decety et al., 1989; Erlacher & Schredl, 2008; Jeannerod, 2001). In the argument, reading books can evoke 'strong representations', which our internal models work on independently of sensory inputs. However, much as we are in the physical world, during a session within virtual reality and film, where continuous input is present, weak representation is active (proximal), as an internal state that bears information about external objects when they are in proximity. According to Clark (Clark, 1997), weak representations are active when the animal is engaged with the world, i.e., the world as its own representation, strong representations become active when the animal is disengaged with the world. As such, the continuous sensorimotor input of virtual reality activates a weak representation as it would in a physical world.

Since presence is considered a perceptual illusion rather than a cognitive one (Slater, 2018), it prioritises basic sensory cues (proximal, or 'weak presentation') as a primary consideration. This happens because the body's automatic perceptual system reacts to an environment more quickly than the conscious, cognitive mind. ISPR (ISPR, 2000) also states that "[a]ll experience of the physical world is mediated by the human senses and complex perceptual processes."

In contrast, a strong representation becomes active when the source object becomes absent (or distal). Imagining (Decety, 1996; Decety et al., 1989), dreaming (Erlacher & Schredl, 2008), and real actions (Jeannerod, 2001) are mediated by the same cortical areas. Neural and cognitive processes are engaged through various activities, including reading, watching a movie, dreaming, and experiencing a virtual environment. Media that are low in immersion, such as photographs, place the "work" of constructing a "spatial situation model on the user or viewer."

An earlier work engages with both weak and strong representations through a process model of spatial presence. Wirth et al. (Wirth et al., 2007) established connections between presence and concepts in psychology and communication. The process model of spatial presence indicated that user-specific factors are crucial, particularly when media technology alone is not adequately immersive to guarantee a sense of presence. These factors operate at different stages of presence formation, influencing a user's attention, mental model construction, and willingness to accept the virtual environment as real. Spatial presence is a two-step process: first, users form a mental model of the media environment by using spatial cues, and second, this mental model becomes the user's primary frame of reference, leading to the experience of being spatially present in that environment rather than their real one.

This body of work leads us to suggest that, with presence as a baseline, the VR community needs to progress further by working towards transport that engages with the cognitive and psychological dimensions through invoking strong representation that works in conjunction with spatial presence. Much like narratives bring readers into storyworlds via a combination of resonances with the protagonist's experiences and spatial cues, thereby evoking mental imagery.

### 2.4 The role of narrative in psychological presence

A body of work in related fields explores the concept of presence beyond VR. For example, presence is termed transport in the study of narratives (R. Gerrig, 2018; Green & Appel, 2024; Green & Brock, 2003), defined as a sequence of events that unfold over time and are causally related to one another (Onega & Landa, 2014a), and the experiences of protagonists (Abbott, 2020) within those stories.

In film theory, it is termed the diegetic effect (Burch, 1979; Tan, 2013), which Burch (p.19) refers to film image as a facsimile of the object, offering a 'perceptual simulation of the real', where spectators experience the diegetic world as an environment. Research has shown that readers can journey into narrative worlds that create distance from their original worlds, and that their beliefs, attitudes, and behaviours are altered by their experience of these narrative worlds (Gerrig, 2023).

This body of research suggests that presence is not solely a function of sensorimotor immersion. Non-spatial presence can emerge from the brain's interpretive and imaginative capacities, independently of technological immersion, as in spatial presence. Stimuli that engage with strong representation, such as narrative, can emotionally and cognitively engage readers, transporting them into storyworlds through resonant details, stimulating imagination, mediated by cortical areas also used in real-world perception and action. Through narratives, readers integrate their memories, personalities, and embodiment into a vivid mental reconstruction of a storyworld. Presence, therefore, is not solely evoked by sensorimotor inputs; narrative in books can engage strong mental activations, internally simulating the world using memory and imagination.

### 2.5 Where do we move on from here?

The sense of 'being there' in another space is a precursor to other experiences. The VR community must progress beyond achieving only spatial presence and understanding its effects. We must work towards creating a more profound sense of transport by designing experiences that also invoke strong representations, and thereby, engaging users on a level beyond the automatic sensory cues of VR to include simulations of other aspects of our world that engage with the cognitive and psychological dimensions. We must move beyond spatial presence by making virtual environments meaningful. To achieve this, designers should look to other fields like storytelling, narrative studies, game design and film, which use similar concepts that balance both weak and strong representations, such as 'transport', 'diegetic effect', to describe how stories can usher users into fictional worlds and even change their beliefs. The ultimate goal is to create experiences that merge automatic sensory cues of VR with the cognitive engagement of narratives. This involves designing systems that leverage both weak representations and strong representations, acknowledging that the final experience is shaped by each user's unique internal mental structures and imagination.

## 3. Dimensionality, Control, Information Flow

A framework is presented here where designers build on the sense of presence, modulating parameters (via information flow) within the dimensions to strategically engage both weak and strong representations.

The technical dimensions create a foundational information space. Our ability to understand and control these technical aspects is the condition that makes this space a viable medium, allowing experiences like social, cultural, placeness, and emotional resonance to emerge.

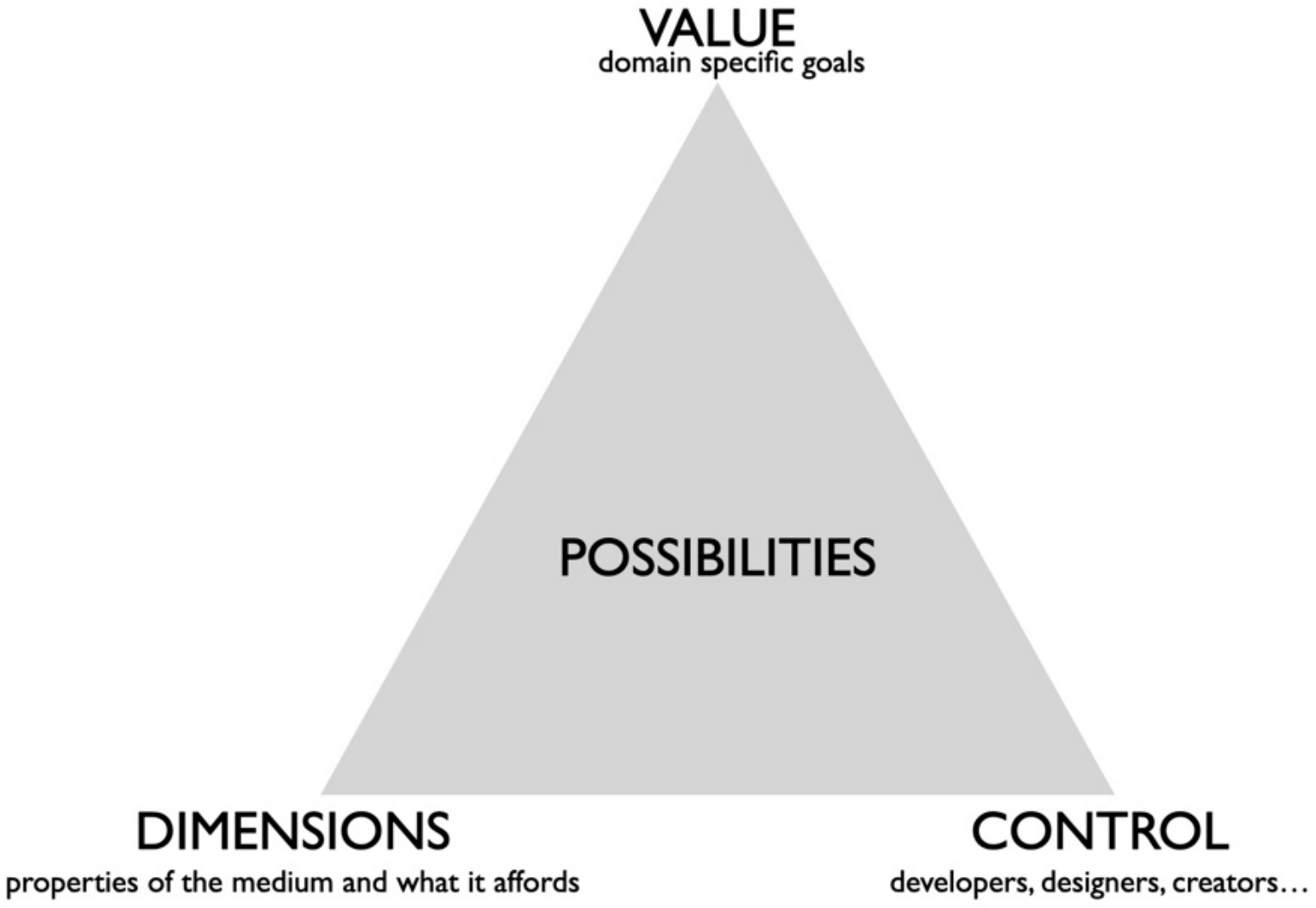

Figure 3. A diagram illustrating the relationship between dimensions, control, possibilities and value. Designers should first define the domain-specific goals, evaluate the dimensions that the properties of the medium provide, and their ability to orchestrate control to achieve the goals.

Our capacity to shape the possibilities of immersive media that engage users in novel ways depends on how we coordinate control. When control is effectively orchestrated by designers through the modulation of properties, the resulting information space can generate possibilities, enabling virtual environments to achieve their purpose and value. Together, our ability to utilize the abstract dimensions will expand the practical potentials (possibilities) of the medium, enabling us to design for a wide range of experiences.

### 3.1 Dimensionality

In this article, "dimensions" refer to modifiable information channels within an immersive environment. Each dimension (spatial, sensory, placeness, temporal, cultural, social, cognitive, psychological) encodes a specific class of information that designers can modulate through defined parameters.

Figure 4 illustrates the concept of the dimensionality spectrum (Cartesian and abstract) that digital technologies present. Most of how we manipulate information is through 2D interfaces. From a purely discrete point of calculation, an additional dimension (z-axis) would certainly add exponential information to the space. Accompanying information, such as textual contents, color (*k* bits per pixel/voxel, e.g., 24-bit RGB → 16.7M colours), degrees-of-freedom (DoF), and interactivity as dynamic states involving time as a 4th dimension, as discrete states, and, if calculated in continuous time, all add to a combination of many possibilities.

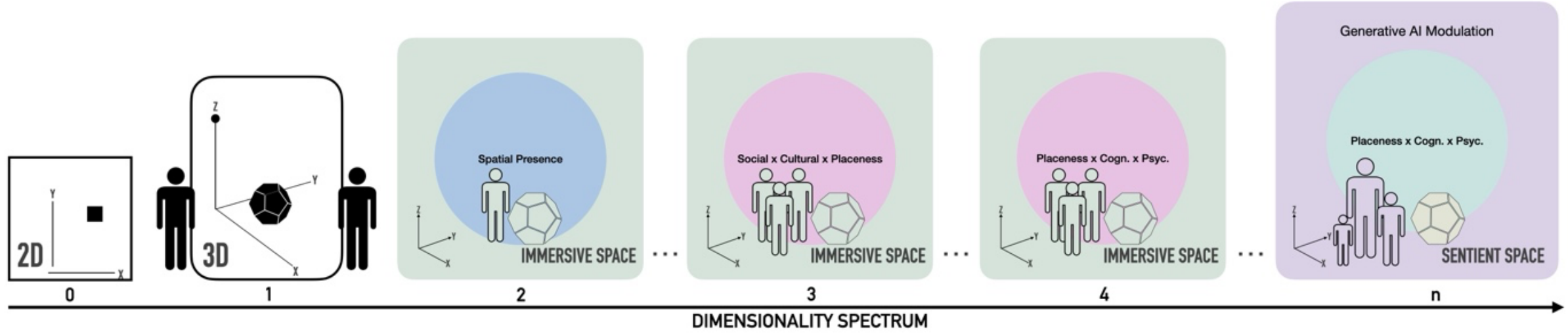

Figure 4. The dimensionality spectrum made possible by technology, from the Cartesian 2D space (0) we are used to, to the extension of the z-axis into a 3D space (1). The sensorimotor immersion (2) giving rise to abstract dimensions (>=3). A dimensionality spectrum would not only describe possibilities, quantified by the information it can encode and decode. It would also allow us to control the information flow within each space. At a certain point in the future, a sentient space appears where designer control is handed over to AI (n).

When, within the Cartesian dimension the sensorimotor component of the immersive media is engaged due to effective control of the Cartesian dimension, spatial presence occurs, when the sensory input system crosses a certain threshold of quality and consistency, following Dooley and Scarbez's argument (Murphy & Skarbez, 2020). Weak representation kicks in, and users are present in the environment, acting and moving naturally. The space itself becomes effective, and when additional abstract dimensions are designed well, it can give rise to a strong representation that encourages users to engage more deeply with the content via the medium. This follows Turner and Turner (Turner & Turner, 2011), the duality of weak (sensory input) and strong representation in the absence of sensorimotor stimulus, and requiring active cognitive processes, such as memory and imagination, and involving higher cognitive functions such as enjoyment (Vorderer & Hartmann, 2009), engagement (Brown & Cairns, 2004; Schoenau-Fog, 2011), involvement (Calleja, 2011; Klimmt & Vorderer, 2003; Vorderer, 1993), absorption (Agarwal & Karahanna, 2000), engrossment (Wilcox-Netepczuk, 2013), immersion (Brown & Cairns, 2004; Calleja, 2011; Ermi & Mäyrä, 2005), and flow (Csikszentmihalyi & Csikzentmihaly, 1990).

Immersive media, in contrast with traditional media (books, TVs), is unique in that it gives rise to a special condition that other media do not afford; it creates a condition for spatial presence as second-order mediation, from which other dimensions can emerge. Users necessarily traverse between weak and strong representations, much like in the real world.

## 3.2 Control

Control in this framework refers to the designer's ability to modulate the informational properties of each dimension (spatial, sensory, placeness, temporal, cultural, social, cognitive, and psychological) to shape the user's experience. It relates to our intention and ability to harness the medium's structures and properties, selectively amplifying, suppressing, or redirecting meaningful information within and across dimensions. By adjusting these parameters ('dials'), designers can create specific experiences that engage with cognitive, emotional, and cultural layers of meaning.

The third dimension is critical in immersive environments because it can evoke spatial presence, which dramatically expands the scope and effectiveness of the information being communicated. This allows designers to encode spatial information in ways that are not possible on a flat, 2D plane. One can imagine the permutations and combinations of elements within a spatial environment such as the evocation of [the appearance] of intensive (temperature, pressure, density, refraction, hardness, etc.) and extensive properties (mass, volume, etc.) that can be simulated and represented within the global and local coordinates (x, y, z) of Cartesian space, communicative meanings in symbols that are associated with the placeness of the environment, which can be culturally dependent (semantics, emotive, functional, historical contexts, aesthetics, etc.) in the abstract dimension, and the changes (temporal dynamics) that occurs as these users interact with these elements and with each other. 3D and immersive environments, therefore, can provide more information than traditional media would allow. They can convey multidimensional, time-dependent, and context-specific information, enabling users to engage with information, events, and social dynamics in ways that are difficult to achieve in traditional media. Given that we can simulate beyond what our physical world is capable of, why are there limits to what one can do, as observed in the many unusable VR applications? I believe that this can be attributed to our ability to control aspects of the medium outside of sensorimotor engagement.

Our capacity to coordinate control over the relational properties of the dimensional spectrum determines how we can make a significant contribution to the entire user experience through the dimensionality spectrum. The amount and timing of information should be modulated in order to prevent cognitive/psychological overload or under-stimulation. The problem of information overload needs to be addressed as users interact with and modify information in these enriched contexts. The key aspect for designers is to create an experience that balances information, enabling meaningful interaction rather than overwhelming complexity.

### 3.3 Information Flow

The flow of information within each scene of the virtual environment is crucial. While present technology has considerably improved upon sensorimotor immersion through advances in the space, earlier studies suggested that the sense of presence is an experience that varies from moment to moment (Heeter, 2003; W. IJsselsteijn et al., 1998; W. A. IJsselsteijn et al., 1997).

Information Flow is how 'meaningful' information (e.g., placeness, social, cultural) is transmitted within the session of a virtual experience. This, of course, goes beyond the technical transmission of information as such (i.e., Shannon and Weaver). Information refers to the transmission of visual, auditory, textual, or haptic signals that are meaningful within spatial and temporal constraints. Information flow refers to the permanence, sequence, pacing, passivity or interactivity (feedback loop), modality, and elaboration of information within the session. The flow control exerted by the designer of an environment ensures that users receive the maximum amount of meaningful information that leads to effects within a given time, without being overloaded or underwhelmed. Well-structured information flows should elicit the intended psychological responses. Information flow could also stimulate conditions for weak (engaged with the immersive environment) and strong representations (engaging with narrative via cognitive processes).

For example, when cognitive load and narrative are minimal, designers could increase interactivity in the virtual environment to engage weak representations (the automatic sense of being there). Conversely, when interactivity is simple, a strong narrative or user cognitive processes could be engaged (strong representation). While the initial feeling of spatial presence is an automatic perceptual illusion of "being there", maintaining a deeper and more meaningful sense of engagement requires this deliberate balance to hold the user's cognitive and psychological focus throughout the experience.

When environments become sentient, the design of the information flow would be supported by AI (ambient intelligence within immersive environments), sensitive and responsive to the state of users. A sentient space leverages generative AI to dynamically alter not just sensorimotor inputs but also narrative and psychological cues based on information learned about its users. In the context of the dimensionality spectrum, imagine bringing a user into an immersive space, a space that is not only a sensory environment, but a virtual world designed to enable a range of cognitive and psychological experiences. Within this deeper immersion, the users' demographics, preferences, and cultural backgrounds become additional information that can enrich the experience. In the not-too-distant future, the informational depth of both physical and virtual spaces will expand as they become increasingly sentient through AI becoming agentic, and with world models, i.e., control of Cartesian and abstract dimensions is passed between designers and AI.

# 4. Dimensionality Framework (8Df)

Eight dimensions are presented in this section: spatial, sensory, placeness, temporal, cultural, social, cognitive, and psychological, as actionable design 'dials' that modulate effects by leveraging the medium's properties and structures, thereby translating theory into practice.

## 4.1 Spatial dimension

The role of the spatial dimension is technical immersion, engaging with weak representation (Turner & Turner, 2006). Many of the challenges within this dimension, such as achieving high-fidelity visualisation and minimising latency, represent long-standing problems in the field of VR that have been solved in recent times.

The agency of VR users is important. Movement along the locomotion spectrum of highly constrained (passive, guided) to completely open (active and degree of freedom), or an interchange of both within zones, affects the users' perceived embodiment (Caserman et al., 2019), control (Al Zayer et al., 2018) and motion sickness (Dużmańska et al., 2018; Kennedy et al., 1993; Kolasinski, 1995; Slater, 2003). There are decades of research in the area, mentioned in section 2. The intended effect is immersion and ultimately the experience of the sense of 'being there' through sensorimotor fidelity.

Table 1. Spatial dimension design: dials, modulations and effects

| | Dial | Modulation | Effects Spectrum |
|---|---|---|---|
| SPA1 | High Perceptual Fidelity<br>vs<br>Low Perceptual Fidelity | High display resolution, wide field of view (FoV), interpupillary distance (IPD) and high-quality spatial audio<br>vs.<br>Low resolution, narrow FoV, fixed IPD, and mono/stereo audio. | A strong sense of being enveloped by the space and high environmental realism<br>or<br>Awareness of the display medium (e.g., screen-door effect, lens-shape, borders, etc.) and a feeling of looking through a viewport with risk of breaks in presence (BIP). |
| SPA2 | Direct 1:1 Mapping<br>vs<br>Altered Mapping | 6-DoF tracking and a room-scale setup that mirrors the physical space<br>vs.<br>Redirected walking, non-isometric scaling, or 3-DoF tracking. | High spatial presence, intuitive interaction, stable sense of place<br>or<br>Ability to navigate larger virtual spaces, but with a potential for disorientation. |
| SPA3 | Natural Locomotion<br>vs<br>Artificial Locomotion | Movement via physical walking or room-scale motion<br>vs. | A stronger sense of embodiment, and reduced motion sickness<br>or<br>Fast and efficient travel through large |
| | | Movement via joystick, teleportation, arm-swinging, or other controller-based methods. | environments, but with a higher risk of cybersickness. |
| SPA4 | Embodied Interaction vs Symbolic/Abstracted Interaction | Isomorphic (one-to-one mapping) interaction via hand tracking, haptic gloves, and direct, physics-based object manipulation vs. Non-isomorphic (scaled linear/non-linear mapping) interaction via controller button presses, ray-casting, and menus. | A high sense of object tangibility and intuitive manipulation or Fast, efficient interaction that may be abstracted, and feel less physical. |
| SPA5 | Full-Body Embodiment vs Disembodied Presence | Fully tracked, animated avatar (via inverse kinematics or body tracking) vs. Third-person/floating camera or simple controller models. | Strong sense of body ownership and embodiment within the VE or Focus on external world with reduced cognitive load and no risk of an "uncanny" avatar. |

## 4.2 Sensory dimension

The sensory dimension is distinct from the spatial dimension in that it facilitates sensorimotor immersion, which induces the sense of presence. The additional sensory channels will only serve to enrich the experience and are tied to the capabilities and limitations of the underlying technology when they become available.

The primary interest is whether multiple sensory stimuli can work cohesively to effectively modulate a variety of moods when such devices have become available. An immersive environment itself that begins with the state of presence as a basis involves multimodality, for example, donning a headset and combining stereoscopic 3D with spatialized audio and haptic feedback, even through micro vibrations in controllers can create a much vivid and believable experience than visuals alone. The degree to which information is aligned and presented across different sensory modalities must be consistent, e.g., a virtual drum synchronising with the visual impact and haptic feedback.

Research has shown that sensory stimulation through the visual, auditory, olfactory, and gustatory systems can modulate mood and alleviate depression (Canbeyli, 2022). Oldfactory and gustatory stimuli are known to affect emotions (Canbeyli, 2022; Dantec et al., 2021; Jabbi et al., 2007). However, research into olfactory simulation within VR is sparse, and studies on gustatory stimuli are even rarer. Olfaction is shown to be emotionally modulated by both personality and mood (D. Chen & Dalton, 2005), which makes smell a powerful factor for achieving emotional resonance, personalisation, and immersion in mediated environments.

Other aspects that simulate other sensory channels could be somatosensation, thermoception, nociception, and equilibrioception (SPA5 of the spatial dimension).

Table 2. Sensory dimension design: dials, modulations and effects

| | Dial | Modulation | Effects Spectrum |
|---|---|---|---|
| SEN1 | Diegetic Sound vs Non-Diegetic Sound | Use of spatialized, synchronized, and environmentally authentic audio cues vs. use of musical scores, UI tones, or voice-over narration. | Deeper immersion and enhanced environmental awareness or Clear guidance, mood induction, and effective communication of information. |
| SEN2 | Contextual Scent vs Scentless Environment | The synchronized delivery of scents that are congruent with the visual context (e.g., the smell of gunpowder during a battle). | Enhanced realism, stronger emotional resonance, and more powerful memory encoding Or |
| | | | avoids scent fatigue or allergies. |
| SEN3 | Contextual Taste<br>vs<br>Tasteless Environment | The delivery of artificial flavor cues that are congruent with the context (e.g., a sweet sensation when eating virtual fruit). | A novel and deeper sensory engagement<br>or<br>avoid hygiene challenges of gustatory simulation. |
| SEN4 | Rich Haptic Feedback<br>vs<br>Simple/No Haptic Feedback | Use of force feedback, texture simulation, and detailed vibrations<br>vs.<br>simple rumble or no feedback. | A strong sense of touch, embodiment, and object tangibility<br>Or<br>a less physically immersive but more accessible and less complex interaction model. |
| SEN5 | Contextual Thermal Cues<br>vs<br>Thermally Neutral Environment | Use of thermoelectric devices to simulate warmth (a fire)<br>vs.<br>cold (a cold breeze) that maps with the visual environment. | Heightened realism and a stronger sense of presence or avoids the technical challenges of thermal simulation. |
| SEN6 | Damage Simulation<br>vs<br>No Damage Feedback | The use of high-intensity, localized haptic jolts (e.g., from a haptic vest)<br>or<br>strong vibrations combined with audiovisual cues to simulate impact. | Increased sense of consequence and risk, creating higher tension and realism<br>or<br>less intense, stressful experience that avoids user discomfort. |

### 4.3 Placeness dimension

The placeness of a location is not the location itself, but it is the center of the human experience, intention, and especially meaning (Relph, 1976). The depth of the meanings that places have for us are informed by the qualities of their settings, referred to as the 'spirit of or identity of the place', or ability to appreciate those qualities of three interwoven elements of a place (Relph, 2007), such as the physical setting, the activities within those settings, and the territories of meaning for an individual or a collective group.

A sense of place can indeed occur within a virtual world. For Relph (ibid.), *insideness* refers to the feeling of being "here" rather than "there", this is a sense of safety, ease, enclosure, and belonging. Its deepest form, *existential insideness*, is the natural, presence experienced in real physical places, such as the feeling of being at home. Relph (2007) also argues that real places are existential phenomena and that virtual places cannot be existential in the same sense. A real sense of place is grounded in lived bodily experience, a virtual sense of place develops through interaction, imagination, immersion, and participation within a digitally mediated environment.

Authenticity, for Relph, involves a direct and genuine experience of a place's identity, emerging organically through long-term everyday life. Because authenticity requires real, lived existence, Relph maintains that virtual places cannot be authentic in the phenomenological sense. A sense of a real place is also synesthetic in that it combines sight, hearing, smell, movement, touch, memory, imagination and anticipation (Relph, 2007). Limited sensory modalities in a virtual environment may not replicate that. However, virtual environments can still be designed to trigger a sense of place by shaping coherent identities, meaningful spatial cues, and opportunities for engagement that allow participants to develop an understanding of virtual insideness, through increasing the weights in other modalities.

Beyond Relph's phenomenology, placeness can arise from the environmental and social characteristics of both physical and virtual settings. The physical or virtual setting itself, as well as the environmental characteristics of the location, the emotions and meanings associated with the environment, activities and social interactions afforded by the place can give rise to placeness (Turner & Turner, 2006). This includes the events, the natural, cultural or social activities, situations, and routines that occur within the place. Individually or collectively (Kyle & Chick, 2007), assigned meanings are attributed to places through ancestral roots, insider status, cultural connection, and personal experiences associated with the place (Hay, 1998). These are more subjective and can be an effect rather than the cause. Measures of the sense of place dwell on social and cultural factors (Shamai & Ilatov, 2005).

Designers may adjust the degree to which a virtual environment resembles the real world or adopts more abstract representations along an abstraction-realism continuum, also referred to as the realism (representation) continuum (Medley & Haddad, 2011). Authenticity, high realism or verisimilitude achieves photorealistic, one-to-one mapping of geometry, lighting, textures, and scale, and is 'indistinguishable from reality' (Slater, 2003). Artificiality involves more abstract, stylised, or cartoonised environments. The designer's goal would be to achieve a functional fidelity, where either a realistic or an abstract environment might be more effective. The temporal dimension of a space, i.e., the amount of time a user spends within a space of a virtual environment or has had experience within it, can contribute to a sense of placeness.

The fidelity of the physical setting also relates to ecological validity, the extent to which a virtual environment supports behaviors or decisions similar to those in real-world contexts (Andrade, 2018; Lewkowicz, 2001). High contextual fit can reduce cognitive load and enhance behavioral plausibility, as in a flight simulator where accurate cockpit affordances support real-world skill transfer. Conversely, in cultural scenarios, such as a reproduction of a traditional South East Asian boat-making scenario for education might not need to reproduce an environment with high fidelity of physical settings, a high level of realism may be unnecessary; placeness may still emerge from selective cues that are meaningful for orientation, narrative, or engagement.

Designers can create the sense of placeness or its opposite effects through the dials in Table 3.

Table 3. Placeness dimension design: dials, modulations and effects

| | Dial | Modulation | Effects Spectrum |
|---|---|---|---|
| PLA1 | Fidelity of physical setting (Place) vs Distortion of physical setting | Spatial layout, routes, connectivity, materials, ecological coherence vs. spatial disarray, pathlessness, fragmentation, immateriality, ecological incoherence. | Recognition, familiarity or displacement, strangeness, curiosity |
| PLA 2 | Inviting Ambience vs Alienating Atmosphere | Emotional tone of space via lighting, sound and atmosphere vs. lacking emotion, dissonance, sterile, ambiguous. | Attachment, nostalgia, joy, comfort or detachment, estrangement, melancholy, unease |
| PLA 3 | Authenticity vs Artificiality | Realism vs. stylization | Sense of reality, verisimilitude or perception of artificiality |
| PLA 4 | Symbolism vs Ambiguity | Icons, metaphors, ancestral reference, cultural symbols vs. The lack of it. | Cultural connection, resonance or misinterpretation and conflict |
| PLA 5 | Activities vs Stasis | Routines, rituals, events, opportunities for social interaction vs. The lack of it. | Reinforcement of place, identity reinforcement, social bonding or dislocation, displacement, and fragmentation |

## 4.4 Temporal dimension

The temporal dimension is linked to the spatial dimension. The ability to simulate virtual time in an effective way that creates temporal compression (NarrativeTime > RealTime) and expansion (NarrativeTime < RealTime) can allow users to psychologically experience narrative time, or fictional time (Juul, 2011), where a user perceives having spent the compressed period in real time. These aspects can be drawn from narrative theory, utilising concepts such as duration, summary, and stretch (Genette, 1980). Transition type can be used to move a user from one point in time or space to another within the narrative of the virtual environment. This certainly affects the sense of continuity and immersion. Transition types can be created to provide a continuous flow, cuts or scene changes, fades, dissolves, or teleportation used in film (Bordwell et al., 2004). If designed well, 'closure' occurs, referring to the mental process of filling in the gaps between the 'comic' panels (McCloud & Martin, 1993), creating a 'continuous unified reality' (p.17).

A virtual environment with high chronological fidelity is one that strictly adheres to a forward-moving sequence of events. Alternatively, time can be manipulated out of chronological order, such as flashbacks (analepsis), flash-forwards (prolepsis) or 'In Medias Res', starting the story in the middle of the action and filling in the backstory later. These are often seen in films and games (Juul, 2011).

Narrative may also branch out in a nonlinear manner, allowing the divergence of events into multiple paths, and in some instances, for events to be replayed with different outcomes (Crawford, 2004). Branching and replayability are linked to user agency and outcomes.

Table 4. Temporal dimension design: dials, modulations and effects

| | Dial | Modulation | Effects Spectrum |
|---|---|---|---|
| TEM1 | Compressed time<br>vs<br>Expanded Time | Alignment of in-world time with real-world physics and user actions<br>vs.<br>time compression (fast-forward) or expansion (slow-motion). | Believability, intuitive interaction, and deep immersion<br>or<br>narrative efficiency, dramatic emphasis, and heightened focus (with a risk of breaking immersion if inconsistent). |
| TEM 2 | Linearity<br>vs<br>Non-Linearity | A single, chronological path<br>vs.<br>non-linear sequences (flashbacks), branching paths, multiple outcomes, and replayability. | A clear, directed, and authored experience<br>or<br>high user agency, exploration, suspense, thematic depth, and replay value (with potential for confusion). |
| TEM 3 | Seamless Transitions<br>vs<br>Abrupt Transitions | Use of continuous flow, dissolves, and fades<br>vs.<br>hard cuts or teleportation. | Sense of continuity, deep immersion<br>or<br>disorientation, narrative jarring, deliberate emphasis. |
| TEM 4 | Implicit Progression<br>vs<br>Explicit Progression | Requiring users to infer or mentally fill temporal gaps<br>vs.<br>explicitly guided progression. | Enhanced engagement through cognitive participation<br>or<br>predictability, guided. |
| TEM 5 | Varied Pacing<br>vs<br>Monotonic Pacing | Rhythm of scenes, mixing moments of high tension with periods of calm.<br>vs.<br>The lack of it. | Sustained engagement, emotional arc, suspense<br>or<br>boredom, lack of emotional impact. |

## 4.5 Cultural dimension

The cultural dimension is closely linked to the placement dimension. It includes the addition of cultural markers (icons, metaphors, and symbols) within the environment's design, such as metaphorical space, aesthetics, and emotional tone. An environment that reflects users' cultural backgrounds, beliefs, aesthetic preferences, or social narratives, and that is familiar, may increase resonance, comprehension, bonding, and emotional engagement. Alternatively, the designer may evoke curiosity, novelty-seeking, induce confusion, or trigger discomfort. Language styles may evoke formality in particular scenarios, and emotional tone may elicit intended moods, such as joy or solemnity.

Figure 5. Cultural dimension design: dials, modulation and effects

| | Dial | Modulation | Effects Spectrum |
|---|---|---|---|
| CUL1 | Culturally Specific<br>vs<br>Culturally Neutral | Use of specific icons, architectural styles, clothing, symbols, and culturally-coded artifacts<br>vs.<br>employing universal symbols. | Deep resonance, strong identification for the target culture<br>or<br>Broad accessibility, wider appeal, potential for feeling generic or placeless. |
| CUL 2 | Cultural Familiarity<br>vs<br>Cultural Novelty | The environment aligns with the user's cultural expectations, norms, and language<br>vs.<br>Introduces foreign or unexpected cultural elements. | Comfort, intuitive understanding, strong sense of belonging<br>or<br>Curiosity, exploration, learning, potential for alienation or confusion. |
| CUL 3 | Embedded Social Norms<br>vs<br>Open Social Norms | Enforcing specific cultural rules for interaction, etiquette, and language (e.g., politeness, proxemics)<br>vs.<br>Freeform, user-defined social space. | Authentic social simulation, clear behavioral guidance<br>or<br>User freedom, emergent social dynamics, potential for cultural clashes. |
| CUL 4 | Explicit Cultural Narrative<br>vs<br>Implicit Cultural Narrative | The environment's stories, rituals, and historical context are clearly presented to the user<br>vs.<br>Cultural context is embedded for users to discover. | Clear contextual grounding, guided cultural learning<br>or<br>Sense of discovery, rewarding exploration, potential for misinterpretation. |

### 4.6 Social dimension

The social dimension refers to how one's virtual representation (Goffman, 2002) and presence relate to other virtual humans and avatars (Bailenson & Blascovich, 2004; Biocca et al., 2003; Gunawardena & Zittle, 1997; Von der Pütten et al., 2010).

The Proteus Effect (Yee & Bailenson, 2007) demonstrates that the appearance of one's avatar can significantly alter the user's own behavior and how others perceive them (Blascovich & Bailenson, 2011). For example, an oversized, disproportionately built avatar may be intimidating, creating distance, whilst an avatar perceived as friendly may attract interaction. The choice between realistic and abstract characters is a key design decision that shapes social interaction (Garau, 2003; Schroeder, 2001a). The social dimension also includes agents.

Alongside human-controlled avatars, some agents are controlled by computers. How designers incorporate avatars and agents within the social dimension is an important consideration, as agents are less persuasive than avatars (Schroeder, 2001b). Learning is improved (Okita et al., 2007), and users are more susceptible to social influence if they believe that they are interacting with a human and not an agent (a machine) (Lim & Reeves, 2010). Further research suggests that perceived avatars produced stronger responses than perceived agents (Fox et al., 2015). It is proposed that social categorisation (Tajfel et al., 2001) may occur (Fox et al., 2015) as in human communities. Interpersonal distance is an important measure (Beall et al., 2003; Roth et al., 2018). Experiments have also shown that participants maintained greater personal space with a human-controlled avatar than with a computer-controlled agent (J. N. Bailenson et al., 2003). This is a vibrant research area of virtual environments.

Table 6. Social dimension design: dials, modulation and effects

| | Dial | Modulation | Effects Spectrum |
|---|---|---|---|
| SC1 | Realistic Avatars<br>vs<br>Stylized Avatars | Realistic reconstructions, natural animations, and human proportions<br>vs.<br>stylized, non-humanoid, or iconic representations. | Stronger self-identification (Proteus Effect), potential for uncanny valley<br>or<br>Focus on role-playing, reduced social pressure, creative expression. |
| SC2 | Human Controlled Avatar<br>vs<br>Complex Agents | World populated with a mix of avatars and believable, socially complex agents<br>vs.<br>Simplified, scripted, or symbolic representations. | Rich self-other recognition, nuanced social cues, identity play, unpredictable bonding, high trust, and a living world<br>vs.<br>Abstracted representation, limited expression, simplified cues, predictable interactions, safe but less dynamic environment, controlled progression |
| SC3 | Emergent Social Norms<br>vs<br>Prescribed Social Norms | Designing open spaces and tools that allow users to establish their own rules for proxemics and interaction<br>vs.<br>Enforcing specific rules for personal space and communication. | User-driven culture, high sense of community ownership<br>or<br>Safe, predictable, and comfortable social experience for newcomers. |
| SC4 | Simulated Social Context<br>vs<br>Abstract Social Context | Creating a realistic environment with authentic social problems and cues to facilitate skill transfer vs.<br>Stylized context focused on gameplay or abstract interaction. | Meaningful learning, high potential for real-world skill transfer<br>or<br>Focus on play, reduced cognitive load, and creative problem-solving. |

## 4.7 Cognitive dimension

The cognitive dimension is a key component of the broader psychological dimension, as our thinking, feeling and behavior are all directly connected. The cognitive dimension is in a separate section, making it distinct in terms of our ability to process information rather than react emotionally.

The cognitive dimension addresses how users process information, avoiding overload, and facilitating mental stimulation (Sweller, 2011, 2020) within multimedia environments (Sorden, 2013). The cognitive process within immersive space is certainly different from focusing one's attention on a 2D display or from reading a book. No doubt, an immersive virtual space can certainly simulate a 2D display and book reading, but spatial presence provides 6 degrees of freedom (6DoF) involving functions in sensory and motor aspects of bodily activity. The attention and memory capacity required for processing information within such environments would significantly differ from, for example, learning from an eBook displayed on a tablet computer, or processing information on how to operate a virtual object, and then navigating from one place to another. The designer's ability to engage users without overwhelming them with redundant information is a crucial component of the cognitive dimension. This necessarily engages with how users process information, including how they learn from information. One of the most prominent metrics that combines the user's cognitive and physical demands is the NASA-TLX (Hart, 2006; Hart & Staveland, 1988). It measures perceived workload across six dimensions: Mental Demand, Physical Demand, Temporal Demand, Performance, Effort, and Frustration. Other aspects are stress and relaxation (Fauveau et al., 2024; Martens et al., 2019), attention (Styles, 2006), memory (Dinh et al., 1999; Mania & Chalmers, 2001; Matheis et al., 2007; Montello et al., 2004), and executive functions, i.e., higher-order cognitive processes such as planning, problem solving, and decision-making (Edsall & Larson, 2006; Goel et al., 2012; Pleban et al., 2001, 2002; Schouten et al., 2010, 2016). Used in VR and AR research, it provides a crucial foundation for what designers must consider in the cognitive dimension.

The complexity of the environment is a crucial variable, as it directly impacts the cognitive load of the user (Godfrey-Smith, 1998; Stokols, 1978). Our ability to navigate the complexities of an environment is a crucial consideration in the design of immersive spaces. Within this Cartesian dimension, designers can control the number of elements, the variety of those elements, their interconnectedness, and their relevance to the environment, all of which directly affect the cognitive load on the user. The aim may be to create a complex environment that increases the user's cognitive load, or another aim may be to induce sensory deprivation. Immersive virtual environment design is not far off from game-level design (Rogers, 2014; Salmond, 2021; Totten, 2019), a well-established area.

An important aspect of our existence within a virtual environment is what we learn from it. Learning engages with cognitive processes. Two important surveys on VR in education are available (Radianti et al., 2020; Wohlgenannt et al., 2019). Learning involves receiving new information that will become useful to continuing in the environment, and also external to it, such as in authentic learning (McClean, 2001), in tangible learning contexts (Pimentel, 1999), and real-life activities (Lombardi & Oblinger, 2007) as opposed to abstract knowledge learned in classrooms. This is especially true when the experience is designed to achieve specific goals, such as a training simulator intended to make one aware of dangers in a real physical setting. Authentic learning in virtual worlds has been discussed earlier (Farley, 2016) in relation to the Second Life virtual world, which criticises the lack of empirical evidence, difficulty in replicating all types of skills, missing non-verbal cues, inauthentic environments, oversimplified problems, lack of complexity, and challenging user interfaces.

Various learning theories can be engaged, these may include more formal methods, such as the cognitive learning theory's 5E model consisting of five stages – engagement, exploration, explanation, elaboration and evaluation (Ruiz-Martín & Bybee, 2022). Knowledge is created through the transformation of experience, and Kolb's theory of experiential learning (Kolb, 1984) is particularly suited to immersive environments. Various researchers have advocated that immersive environments can accommodate experiential learning (Bell & Fogler, 1997; C. J. Chen et al., 2005; Fromm et al., 2021; Majgaard & Weitze, 2020; San Chee, 2001). Kolb's experiential learning states that students cycle through the four different learning modes of concrete experience, reflective observation, abstract conceptualization, and active experimentation. Learning is a process rather than an outcome. Immersive virtual environments can certainly accommodate the preferred ways of processing and internalizing information, which were identified by Kolb's four learning styles – diverging, assimilating, converging, and accommodating (Fromm et al., 2021; Majgaard & Weitze, 2020). Understanding user behaviour within immersive environments then becomes necessary if we want to communicate information effectively (Cai et al., 2018; Ch'ng, Li, et al., 2019; Li et al., 2021).

Table 7. Cognitive dimension design: dials, modulation and effects

| | Dial | Modulation | Effects Spectrum |
|---|---|---|---|
| COG1 | Cognitive Overload<br>vs<br>Cognitive Underload | High amount, complexity, and rapid pacing of information<br>vs.<br>Low amount, simplicity, and slow pacing. | Cognitive fatigue, confusion, and anxiety<br>or<br>Boredom and disengagement. |
| COG2 | Directed Attention<br>vs<br>Ambient Awareness | Clear goals, problem-solving tasks, and minimal information load<br>vs.<br>Open-ended exploration, rich background details, and non-critical events. | High mental engagement, skill development, a sense of "flow"<br>or<br>Relaxation, mindfulness, a sense of awareness without pressure. |
| COG3 | Experiential Learning<br>vs<br>Explicit Instruction | Use of diegetic interfaces and discovery-based tasks (e.g., Kolb's cycle)<br>vs. | Deep immersion, strong knowledge retention through experience |
| | | Non-diegetic HUDs, tutorials, and structured guidance (e.g.,5E model). | or<br>High degree of control, clarity, and efficient, structured learning. |
| COG4 | Rote Memorization<br>vs<br>Contextual Learning | Use of repetition and decontextualized drills<br>vs.<br>Embedding knowledge within spatial cues, experiential tasks, and meaningful narratives. | Efficient encoding of specific facts<br>or<br>Deeper understanding and better real-world skill transfer. |
| COG5 | User-Directed Agency<br>vs<br>System-Directed Agency | Allowing the user to set their own goals, make meaningful decisions, and control the pace<br>vs.<br>Guiding the user with linear paths and system-initiated events. | Sense of freedom, ownership, intrinsic motivation<br>or<br>Clarity of purpose, reduced decision fatigue, a curated and guided experience. |

### 4.8 Psychological dimension

Emotion is evolution's way of giving meaning to our lives (Bower, 2014), perhaps within virtual experiences too. Psychological effects, particularly positive emotions, are a necessary dimension in the design of virtual experiences. Immersive environments are well-suited for learning that is linked to emotions and moods. This section provides some guidelines for engaging with the user's emotions.

The cognitive dimension is a part of the psychological dimension. Our thinking, feeling and behavior are all directly connected (Nabavi & Bijandi, 2012). Cognitive processes may lead to psychological effects. There are many types of psychological constructs measured in VR to understand the medium's effects on users. These include emotional arousal, emotional valence, dominance, and specific emotions such as joy, sadness, anger, fear, anxiety, boredom, hopelessness, and surprise.

One of the greatest drivers of these psychological effects is the narrative and story told through the virtual environment. VR technology can create a high-fidelity reality simulator, but while the technical properties contribute to the believability of the immersive space, it is the content of the narrative that directs and produces specific psychological outcomes such as fear, anxiety, joy, sadness or surprise.

A comparative analysis of major theoretical frameworks reveals a convergence around several core components that consistently underpin effective storytelling across traditions. While individual scholars emphasize different aspects, the literature collectively highlights the centrality of characters or agents (Halliwell, 1998; Herman, 2004, 2009; Ryan, 2015b; Truby, 2008), the structuring of plot and events (Campbell, 2008; Chatman & Chatman, 1978; Halliwell, 1998; Truby, 2008), and the articulation of themes and meaning (Fisher, 1984, 2021; Halliwell, 1998; Ricoeur & Ricoeur, 1984). Complementary to these are the significance of the setting or storyworld (Chatman & Chatman, 1978; Herman, 2004; Ryan, 2015b), the shaping of narrative through discourse and medium (Chatman & Chatman, 1978; Ricoeur & Ricoeur, 1984; Ryan, 2015b), and the availability of transformation or resolution within the story arc (Bucher, 2017; Campbell, 2008; Truby, 2008). Finally, effective storytelling is marked by its capacity to achieve resonance with audiences, ensuring coherence and fidelity across cultural and experiential contexts (Fisher, 1984, 2021; Ricoeur & Ricoeur, 1984; Ryan, 2015b).

Most virtual environments are descriptive, i.e., they present states and describe them (Onega & Landa, 2014b, p. 5), when they can be much more. Instead, virtual environments can provide a narrative and deliver it via storytelling, emphasising an interpretation for different audiences using the properties of the medium. If a virtual environment is descriptive, it is not using the full dimensions of the medium it affords (see section 3).

According to neuroscientific research, emotions are deep biological processes that help us assign value and guide survival-oriented actions (Panksepp, 2004). Positive emotions such as joy and curiosity contribute to learning. Positive feelings are associated with the release of dopamine, which plays a crucial role in motivation, memory and learning (Wise, 2004). In self-directed learning, for example, dopamine release locks attention on the item, task, or concept currently being attended to or represented (Herd et al., 2010). When the subject is enjoyable, learners are motivated to engage with it, their attention is directed (Tyng et al., 2017), and they spend more time exploring the subject in depth. Emotions are deeply intertwined with the cognitive processes of attention, memory, and motivation, acting as a powerful amplifier or inhibitor of learning.

This deep link between emotion and cognition is especially relevant for learning in virtual environments (Pekrun, 1992, p.364-369). In contrast to traditional media, donning a headset and entering into another world is certainly different from classroom learning or exam halls. It feels more like play, and therefore, positive emotions might be experienced more frequently within such environments. Therefore, the absence of positive emotions can be a 'dial' from which specific effects are created. Negative emotions can be leveraged to create a lasting impression. While fear, anxiety, panic, and other similar emotions can inhibit learning, they can also be leveraged to create urgency in situations that require it.

The unique power to direct a user's emotional state applies not only to learning; it extends to facilitating deeper social-moral experiences like empathy that lead to changes in behavior. Aylett and Louchart (Aylett & Louchart, 2003) considered VR a particular narrative medium alongside theatre, literature or cinema. Filmmaker Chris Milk, in a TED talk delivered in 2015, heralded VR as "the ultimate empathy machine." VR can immerse users in a lived experience of others, creating emotional impact and resonance, especially through embodiment. It can be hypothesised that stories in this medium would have more of an impact on the transformation and impact on its active participants (as opposed to the audience). Stories can be designed to change perceptions or behavior. One of the levers designers could use might be to integrate actions within emotional storytelling that make use of the social-moral compass (Haidt, 2003), and thus create a personal transformation for the users.

Bucher (Bucher, 2017) highlighted several key differences in VR and traditional media such as film, television, or literature. The distinctions revolve around similar themes. These are immersion, agency, and interactivity. Audiences are observers in traditional media, experiencing the virtual world from a fixed perspective (camera, narrator, or page). In VR, users are participants inhabiting the virtual world. The multisensory environment surrounds them, and users influence the pacing, alter the outcomes, and the experience branches. Time is not sequenced and in a fixed order. Spaces can be revisited, and events are experienced in different orders.

As this is an exploratory area that is in its infancy, a new grammar of narrative based on attention cues, spatial audio, gaze direction, and interactivity as opposed to edits and cuts in traditional storytelling media can be explored. There is no single best practice as such, and practices must be drawn from theatre and games, more than cinema and TV. Bucher's deep 'Spotlight' discussions with experts in the field are a good start (Bucher, 2017).

Table 8. Psychological dimension design: dials, modulation and effects

| | Dial | Modulation | Effects Spectrum |
|---|---|---|---|
| PSY1 | Positive Valence<br>vs<br>Negative Valence | Use of joyful aesthetics, audio and rewarding feedback<br>vs.<br>Urgent tasks, suspenseful audio, and dissonant visuals. | Feelings of joy, comfort, and safety or Feelings of fear, anxiety, and urgency. |
| PSY2 | Active Protagonist<br>vs | Branching narratives where user choices have consequences | A high sense of agency, responsibility, and narrative ownership |
| | Passive Observer | vs.<br>Linear, non-interactive story with a predetermined user path. | or<br>Curated, relaxing, and guided narrative experience. |
| PSY3 | Intrinsic Motivation<br>vs<br>Extrinsic Motivation | Supports curiosity with open worlds and discovery-based goals<br>vs.<br>Using explicit rewards, points, and gamified objectives (extrinsic rewards). | Deep, sustained engagement driven by curiosity and interest<br>or<br>Clear, goal-oriented focus driven by rewards and progression. |
| PSY4 | Predictable Events<br>vs<br>Unpredictable Events | Consistent and expected event progression and feedback<br>vs.<br>Use of surprise, unexpected events, and novel stimuli. | A sense of safety and comfort, efficient learning of patterns<br>or<br>Heightened attention, stronger memory encoding, and excitement (with risk of anxiety). |

## 4.9 Using the framework in a qualitative evaluation

In recent years, elaborate VR applications that depict cultural content via strong narratives and storytelling approaches have begun to populate exhibition spaces. These VR experiences utilized large, open spaces that accommodated a multitude of users, grouped into 4-6 users per tour. A group of 50 students and instructors attended the exhibition. They form groups of 3-4 partners per session, from which we recorded their experience through verbal accounts.

One such is the *Horizon of Khufu: A Journey in Ancient Egypt*, a 45-minute VR experience that we attempted to fit within the dimensionality spectrum. The experience within a physical interactive space of 10,000 to 11,000 square feet has several key scenes that define the experience, in the following order:

1. A starting scene with Mona as the tour guide outside the pyramids.
2. An exploration of the interior of the Pyramids, moving through tunnels and chambers that were normally closed to the public, such as the Grand Gallery and King's Chamber.
3. A brief encounter with the goddess Bastet, transformed from a house cat that accompanied us in the inner chambers, after the cat knocked the torch onto the ground.
4. An ancient boat ride on the reconstruction of the Nile River.
5. An embalming ceremony and funeral rites for King Khufu, with priests and mourners.
6. An ascent to the top of the Great Pyramid for a panoramic view of the Giza plateau.
7. The user becomes a giant in a scene that explores the Giza plateau.
8. A goodbye scene with a summary by Mona, the tour guide.

The Horizon of Khufu underwent improvements in hardware, e.g., the removal of the computer from the backpack, from my first visit to the second visit over a two-year period. The VR experience was purposeful, contextually relevant, educational and entertaining. The VR experience has sufficient depth and scope that can be qualitatively evaluated, using all the properties of the dimensions proposed in section 3.

The technical and spatial dimensions certainly scored highly. There was a 1:1 mapping in all scenes, except for the scene where users became giants, towering above the pyramids. Even with altered mapping, presence was felt. Locomotion felt natural, and there was a sense of embodiment as we assumed the simple avatar as our embodiment. Our gaze, position, rotation, limbs, gaze and fingers were tracked. There was a strong sense of body-ownership and virtual objects, including walls and ceilings of narrow passages. They felt physical even with the stylized rendering. Many of us were seen stooping on scenes with low ceilings and crawling as we attempted to get through an enclosure, reacting naturally as if the space was physical.

The spatial design of the journey was optimized for storytelling, without redundancies such as extraneous spaces. Spaces were sufficiently complex; the large expanse of the landscape and the claustrophobic depiction of the pyramid's interiors were realistic to users of different sizes and heights.

The spatial design was bounded, and time spent within the space was brief but adequate (temporal dimension), and yet there was no sense that the agency of the users was restricted. Most VR experiences were capped at 10-20 minutes from user preferences and experience (Ch'ng, Cai, et al., 2019; Ch'ng, Li, et al., 2019), but the 45-minute journey did not feel lengthy. There was real-world time used in each session, except for when transitions occur that transport us to a different scene. The change between scenes was a combination of seamless transitions, and others were abrupt, i.e., via a transitional virtual space emphasizing that we are moving into a different scene. Time was linear, and explicit progression was applied, resulting in a very low cognitive load.

The sensory (sensory dimension) inputs were fitting for the type of educational storytelling of the virtual environment, even though the visual quality of the textured models and scenes was not high-quality. For example, visual fidelity, defined as 'the extent to which the VE and interactions with it are indistinguishable from the participant's observations of and interactions with a real environment.' (Waller et al., 1998), would be how information is presented, perceived and felt. Even though the Horizon of Khufu provides mid-ranging visual fidelity (according to an expert 3D modelling instructor in our school), the other aspects of information, such as storytelling that engages with cognitive and psychological dimensions, would balance or make up the total experience of a user. This indicated that the components of the dimensionality framework can be used interchangeably.

Many of the visitors have not been to the Pyramids; they have only seen them in photos and videos. The spatial layout, routes, and materials were therefore unfamiliar, and a certain strangeness and curiosity were part of the experience (Placeness Dimension). However, the design of the lighting, sound and warm hues of the environment was not alienating; instead, it was inviting and comforting. The environment and objects were not rendered as realistic representations, and this mid-point visual fidelity between realism and cartoonish rendering created a fictional experience. There was an instance of ambiguity before we entered the tunnels. This was the case of the security guard at the entrance, who was out of place. Overall, icons, metaphors and cultural symbols were definite, and recognizable. The entire journey required full participation, for one needed to move toward the next scene through bodily activity, e.g., walking, crawling, or moving around a subject to inspect it. Activities of the participants, however, were as observers and learners, without being involved in any of the rituals.

The environment is also culturally meaningful (placeness and cultural dimension), detailed and optimized in its use of symbols, e.g., clothing, Bastet, the Egyptian goddess, a guide to the dead serving as a protector of homes and families, and associated with fertility, childbirth, and women's secrets, as well as the afterlife. The experience also depicted metaphorical space (e.g., the spatial as temporal – travel through time, transition between life and afterlife, the Pyramid as a portal), and emotional tones (e.g., at the funeral rites).

The social dimension was between the users themselves, as they conversed during the journey, and between Mona, the tour guide, and the Egyptian cat deity. Accompanying our journey was our acquaintances (Four human-controlled avatars in my group, three in others), users' voices and avatars (mid-realism fidelity), whom we could bump into when in proximity. The avatar was a semi-transparent white figure with particle effects, gaze, position, rotation, limbs, gaze and fingers were tracked (spatial dimension). After just a few minutes in the scene, one seems to have adopted our own avatar. The acquaintances being in the same space provided a sense of safety in an unfamiliar environment, and the modulation of Mona's voice, the friendly-looking agent as a guide, was to create a sense of trust and safety.

There were simple scripted agents, the guard at the entrance, Mona, the cat which became Bastet, and the priests and mourners at the funeral. The user reaction to these was safe and predictable interactions. Prescribed social norms were implemented, and thus, a predictable, comfortable social experience was created. Social contexts were not implemented in this case.

Although engagement with Mona and the Egyptian cat deity was monodirectional, it felt as if it were bidirectional. Both Mona and the cat deity guided us across the various scenes of the journey. The design was well-conceived, and the overall feeling was that there had been real interaction and engagement.

In terms of the cognitive dimension, the experience engages users by merging spatial navigation with historical reconstruction, and these stimulate the thinking process as users synthesize sensory input, cultural context, and temporal dislocation into a coherent narrative of a part of ancient life at the Pyramids. The spatial, placeness, and temporal dimensions were very well modulated, which considerably reduces cognitive processing. In the experience, the psychological dimensions were engaged more than the cognitive dimensions.

Psychologically, users experienced a strong emotional resonance throughout the VR journey. Many of our students expressed a sense of affective engagement with the virtual tour guide, Mona, noting feelings of safety and companionship. Some of us even remarked on their reluctance to leave ancient Egypt, as if we were actually there. This highlights the social, cultural experience and the emotional bonds formed within the experience.

In summary, the placeness, social, cultural, cognitive, and psychological dimensions have superseded the intermediate quality of computer graphics rendering. The balance of how information flows and control was orchestrated between the dimensions, in particular, the guided structure of the narrative, the directed but spatial freedom of the scenes, and the emotional engagement provided by the tour guide's storytelling created the possibility for such an experience. At the apex of the experience of such a possibility lies value: the value of a deeper understanding of ancient culture within a historical context, a memorable personal connection with other users and avatars, and a sense of having lived through history rather than merely reading a book or watching a documentary.

Table 9. Proof of concept mapping of the theoretical framework, incorporating the analysis of the Horizon of Khufu experience. It is important to note that the balance of the use of the 'dials' is important, and that not all 'dials' are necessary for a good experience.

| Dimension | Dial | Modulation in Horizon of Khufu | Effects Spectrum |
|---|---|---|---|
| Spatial | SPA1: High Perceptual Fidelity vs. Low Perceptual Fidelity | Mid-ranging visual fidelity with stylized rendering rather than photorealism. | An awareness of the display medium was present, but this was balanced by strong storytelling to maintain the total experience. |
| | SPA2: Direct 1:1 Mapping vs. Altered Mapping | Primarily 1:1 mapping was used, with one scene featuring altered mapping where users became giants. | A high sense of spatial presence was felt even with the altered mapping. |
| | SPA3: Natural Locomotion vs. Artificial Locomotion | Locomotion was natural, achieved through physical walking, tracked limbs and fingers. | This resulted in a stronger sense of embodiment. |
| | SPA4: Embodied Interaction vs. Symbolic/Abstracted Interaction | Interaction was embodied; users felt the physicality of virtual objects and were observed stooping under low ceilings. | This created a high sense of object tangibility and intuitive manipulation. |
| | SPA5: Full-Body Embodiment vs. Disembodied Presence | Users were represented by a simple, tracked avatar. | A strong sense of body-ownership was achieved as users quickly adopted their avatars. |
| Sensory | SEN1: Diegetic Sound vs. Non-Diegetic Sound | Utilized non-diegetic sound, such as the voice of Mona, the tour guide. | This provided clear guidance and induced a mood of trust and safety. |
| Placeness | PLA1: Fidelity of physical setting vs. Distortion of physical setting | For many visitors, the setting was unfamiliar as they had not been to the Pyramids. | This created an experience of strangeness and curiosity. |
| | PLA2: Inviting Ambience vs. Alienating Atmosphere | The lighting, sound, and warm hues were designed to be inviting and comforting. | This resulted in feelings of attachment and comfort. |
| | PLA3: Authenticity vs. Artificiality | The environment used a mid-point visual fidelity between realism and cartoonish rendering. | This created a perception of a fictional experience rather than a sense of reality. |
| | PLA4: Symbolism vs. Ambiguity | The experience was definite and recognizable in its use of icons, metaphors, and cultural symbols, with one minor instance of ambiguity (a security guard). | This resulted in cultural connection and resonance, even though the participants had not been to Egypt before. |
| | PLA5: Activities vs. Stasis | The journey required full participation via physical movement, though users were primarily observers rather than participants in rituals. | This reinforced place identity. |
| Temporal | T1: Compressed time vs. Expanded Time | The experience mostly used real-world time in each session. The 45-minute journey did not feel lengthy. | This contributed to believability and intuitive interaction. |
| | T2: Linearity vs. Non-Linearity | The narrative followed a single, linear path. | This provided a clear, directed, and authored experience, reducing cognitive load. |
| | T3: Seamless Transitions vs. Abrupt Transitions | A combination of seamless and abrupt transitions were used between scenes. | The abrupt transitions deliberately emphasized movement into a different scene. |
| | T4: Implicit Progression vs. Explicit Progression | Explicit progression was applied throughout the experience. | This resulted in predictability and potential spoon-feeding, reducing cognitive load. |
| | T5: Varied Pacing vs. Monotonic Pacing | The 45-minute session did not feel long, implying a varied and engaging pace. | This sustained engagement and maintained an emotional arc. |
| Cultural | C1: Culturally Specific vs. Culturally Neutral | The environment was detailed with specific cultural symbols such as clothing and deities. | This created deep resonance and a strong identification for the target culture. |
| | C2: Cultural Familiarity vs. Cultural Novelty | For most users, the environment introduced foreign and unexpected cultural elements. | This sparked curiosity, exploration, and learning. |
| | C3: Embedded Social Norms vs. Open Social Norms | Prescribed social norms were implemented. | This resulted in a safe, predictable, and comfortable social experience. |
| | C4: Explicit Cultural Narrative vs. Implicit Cultural Narrative | The cultural context was clearly presented through the narrative and tour guide. | This provided clear contextual grounding and guided cultural learning. |
| Social | SC1: Realistic Avatars vs. Stylized Avatars | Avatars were stylized as semi-transparent white figures with particle effects. | This stylized choice allowed users to quickly adopt the avatar and feel embodied without the risk of an "uncanny" effect. |
| | SC2: Human Controlled Avatar vs. Complex Agents | The world contained a mix of human-controlled avatars and simple, scripted agents (Mona, priests). | The presence of human avatars provided a sense of safety, while agents offered predictable interactions that felt engaging. |
| | SC3: Emergent Social Norms vs. Prescribed Social Norms | The experience used prescribed social norms. | This created a safe and comfortable social experience for newcomers. |
| | SC4: Simulated Social Context vs. Abstract Social Context | Social contexts were not a primary focus of the experience. | The focus was on the historical narrative rather than meaningful learning for real-world skill transfer. |
| Cognitive | COG1: Cognitive Overload vs. Cognitive Underload | The experience was designed with a very low cognitive load. | This allowed for a relaxed and engaging experience, avoiding confusion or fatigue. |
| | COG2: Directed Attention vs. Ambient Awareness | The experience stimulated the thinking process by merging spatial navigation with historical reconstruction, requiring users to synthesize information. | This resulted in high mental engagement and a coherent narrative understanding. |
| | COG3: Experiential Learning vs. Explicit Instruction | The journey was a form of experiential learning where users learned through active participation and discovery, as observers. | This led to a deeper understanding and strong knowledge retention through experience. |
| | COG4: Rote Memorization vs. Contextual Learning | Knowledge was embedded within the spatial cues and narrative of the experience. | This facilitated deeper understanding and better real-world skill transfer compared to decontextualized drills. |
| | COG5: User-Directed Agency vs. System-Directed Agency | The narrative was system-directed, guiding the user along a linear path, although users had freedom of movement within scenes. | This provided clarity of purpose and a curated, guided experience. |
| Psychological | PSY1: Positive Valence vs. Negative Valence | Users experienced strong positive emotions, including feelings of safety, companionship, and emotional resonance. | This resulted in feelings of joy, comfort, and safety. |
| | PSY2: Active Protagonist vs. Passive Observer | Users were active participants in the journey but passive observers of the linear, non-interactive story. | This provided a curated, relaxing, and guided narrative experience. |
| | PSY3: Intrinsic Motivation vs. Extrinsic Motivation | Engagement was driven by curiosity and interest in the open-ended exploration of a historical world. | This led to deep, sustained engagement driven by curiosity and interest. |
| | PSY4: Predictable Events vs. Unpredictable Events | The progression of events, guided by the tour guide, was consistent and expected. | This created a sense of safety and comfort. |

# 5. Conclusion

This article presents a new framework for designing and evaluating virtual reality experiences that advocates moving beyond the traditional focus on achieving 'presence'. The article argues that 'being there' should be considered a baseline expectation for VR, and not the ultimate objective. The focus should instead shift to orchestrating a variety of abstract dimensions afforded by the spatial dimension to create specific, meaningful, and valuable user experiences.

The article begins by contextualising the historical focus of VR research on technology and systems designed to enhance immersion. As technology progresses, this focus is no longer sufficient for evaluating the depth of immersive media. Many VR applications fail to deliver meaningful experiences because they underutilize the medium's potentials, often by simply replicating physical environments without a clear purpose beyond looking real. The article contends that once a user feels present in a virtual space, even an empty one, the immediate question becomes, "Why am I here, and what can I do?". This highlights the present need to prioritise the 'content' of the experience once the 'form' is achieved.

The core of the contribution is a new framework built on the relationship between the three key concepts: dimensions, control, and value, which intersect to create possibilities in the medium. The framework introduces a dimensionality spectrum that designers can manipulate to achieve the desired effect, beginning with the foundational technical dimension that enables spatial presence and builds upon it with several abstract dimensions to engage users cognitively and psychologically. For each of the technical (spatial, sensory) and abstract dimensions (placeness, temporal, social, cultural, cognitive, psychological), a series of 'dials' is proposed that can be modulated to achieve a spectrum of effects. These dimensions are tied together using the concept of information flow, which refers to how meaningful information is orchestrated and transmitted over time to guide the users' journey. Effective control of this flow prevents cognitive load and balances the users' engagement between automatic sensory cues (weak representation) and deeper cognitive involvement (strong representation).

Finally, the paper demonstrates the framework's utility by applying it in a qualitative evaluation of the VR experience, using the Horizon of Khufu: A Journey in Ancient Egypt as a case example.

There are still unknowns in the orchestration of the properties of the dimensions, and these are the way forward for research in the community. As a community, our research should move beyond technical development, shifting toward a focus on how to design for the experience. This shift in theoretical focus would open up new avenues of investigation into the quality of the human experience, focusing on the rich, subjective, and affective dimensions of cognitive and psychological experience within immersive environments. This necessary transformation invites deeper interdisciplinary dialogue from other communities, drawing on philosophy, cognitive science, affective theory, media and cultural studies, and the arts to gain a more comprehensive understanding of how immersive technologies shape, and are shaped by, human consciousness and culture.

**References**

Abbott, H. P. (2020). *The Cambridge introduction to narrative*. Cambridge University Press.

Agarwal, R., & Karahanna, E. (2000). Time flies when you're having fun: Cognitive absorption and beliefs about information technology usage. *MIS Quarterly*, 665–694.

Al Zayer, M., MacNeilage, P., & Folmer, E. (2018). Virtual locomotion: a survey. *IEEE Transactions on Visualization and Computer Graphics*, *26*(6), 2315–2334.

Andrade, C. (2018). Internal, external, and ecological validity in research design, conduct, and evaluation. *Indian Journal of Psychological Medicine*, *40*(5), 498–499.

Aylett, R., & Louchart, S. (2003). Towards a narrative theory of virtual reality. *Virtual Reality*, *7*(1), 2–9.

Bailenson, J. N., & Blascovich, J. (2004). Avatars. *Encyclopedia of Human-Computer Interaction*, *1*, 64–66.

Bailenson, J. N., Blascovich, J., Beall, A. C., & Loomis, J. M. (2003). Interpersonal distance in immersive virtual environments. *Personality and Social Psychology Bulletin*, *29*(7), 819–833.

Bailenson, J., Yee, N., & Blascovich, J. (2008). Guadagno Rosanna E. 2008. Transformed social interaction in mediated interpersonal communication. In *Mediated interpersonal communication.* (pp. 77–99). Routledge.

Barfield, W., & Weghorst, S. (1993). The sense of presence within virtual environments: A conceptual framework. *Advances in Human Factors Ergonomics*, *19*, 699.

Barfield, W., Zeltzer, D., Sheridan, T., & Slater, M. (1995). Presence and performance within virtual environments. *Virtual Environments and Advanced Interface Design*, 473–513.

Beall, A. C., Bailenson, J. N., Loomis, J., Blascovich, J., & Rex, C. S. (2003). Non-zero-sum gaze in immersive virtual environments. *Proceedings of HCI International*.

Bell, J. T., & Fogler, H. S. (1997). Ten steps to developing virtual reality applications for engineering education. *1997 Annual Conference*, 2–402.

Biocca, F. (1999). The cyborg's dilemma: Progressive embodiment in virtual environments. *Human Factors in Information Technology*, *13*, 113–144.

Biocca, F. (2003). Can we resolve the book, the physical reality, and the dream state problems? From the two-pole to a three-pole model of shifts in presence. *EU Future and Emerging Technologies, Presence Initiative Meeting*, *12*, 13–69.

Biocca, F., Harms, C., & Burgoon, J. K. (2003). Toward a more robust theory and measure of social presence: Review and suggested criteria. *Presence*, *12*(5), 456–480.

Biocca, F., & Levy, M. R. (2013). *Communication in the age of virtual reality*. Routledge.

Blascovich, J. (2002). Social influence within immersive virtual environments. In *The social life of avatars: Presence and interaction in shared virtual environments* (pp. 127–145). Springer.

Blascovich, J., & Bailenson, J. (2011). *Infinite reality: Avatars, eternal life, new worlds, and the dawn of the virtual revolution*. William Morrow & Co.

Bordwell, D., Thompson, K., & Smith, J. (2004). *Film art: An introduction* (Vol. 7). McGraw-Hill New York.

Bower, G. H. (2014). How might emotions affect learning? In *The handbook of emotion and memory* (pp. 3–31). Psychology Press.

Brooks Jr, F. P. (1999). What's real about virtual reality? *Computer Graphics and Applications, IEEE*, *19*(6), 16–27.

Brown, E., & Cairns, P. (2004). A grounded investigation of game immersion. *CHI'04 Extended Abstracts on Human Factors in Computing Systems*, 1297–1300.

Bucher, J. (2017). *Storytelling for virtual reality: Methods and principles for crafting immersive narratives*. Routledge.

Burch, N. (1979). *To the distant observer: Form and meaning in the Japanese cinema*. Univ of California Press.

Burdea, G. C., & Coiffet, P. (2003). *Virtual reality technology*. John Wiley & Sons.

Cai, S., Ch'ng, E., & Li, Y. (2018). A Comparison of the Capacities of VR and 360-Degree Video for Coordinating Memory in the Experience of Cultural Heritage. *Digital Heritage 2018*.

Calleja, G. (2011). *In-game: From immersion to incorporation*. mit Press.

Camci, A. (2019). Exploring the effects of diegetic and non-diegetic audiovisual cues on decision-making in virtual reality. *Paper/Exploring-the-Effects-of-Diegetic-and-Non-Diegetic-% C3% 87amci/E3bf1083e493ba68b81d47e45f1d5295e17302d4*.

Campbell, J. (2008). *The hero with a thousand faces* (Vol. 17). New World Library.

Canbeyli, R. (2022). Sensory stimulation via the visual, auditory, olfactory and gustatory systems can modulate mood and depression. *European Journal of Neuroscience*, *55*(1), 244–263.

Caserman, P., Garcia-Agundez, A., & Göbel, S. (2019). A survey of full-body motion reconstruction in immersive virtual reality applications. *IEEE Transactions on Visualization and Computer Graphics*, *26*(10), 3089–3108.

Chatman, S. B., & Chatman, S. (1978). *Story and discourse: Narrative structure in fiction and film*. Cornell university press.

Chen, C. J., Toh, S. C., & Ismail, W. M. F. W. (2005). Are learning styles relevant to virtual reality? *Journal of Research on Technology in Education*, *38*(2), 123–141.

Chen, D., & Dalton, P. (2005). The effect of emotion and personality on olfactory perception. *Chemical Senses*, *30*(4), 345–351.

Ch'ng, E., Cai, S., Leow, F.-T., & Zhang, T. (2019). Adoption and use of emerging cultural technologies in China's museums. *Journal of Cultural Heritage*, *37*. https://doi.org/10.1016/j.culher.2018.11.016

Ch'ng, E., Li, Y., Cai, S., & Leow, F.-T. (2019). The Effects of VR Environments on the Acceptance, Experience and Expectations of Cultural Heritage Learning. *Journal of Computing and Cultural Heritage*.

Choi, S., Jang, C., Lanman, D., & Wetzstein, G. (2025). Synthetic aperture waveguide holography for compact mixed-reality displays with large étendue. *Nature Photonics*, 1–10.

Clark, A. (1997). The dynamical challenge. *Cognitive Science*, *21*(4), 461–481.

Clark, A. (1998a). *Being there: Putting brain, body, and world together again*. MIT press.

Clark, A. (1998b). *Being there: Putting brain, body, and world together again*. MIT press.

Clark, A., & Grush, R. (1999). Towards a cognitive robotics. *Adaptive Behavior*, *7*(1), 5–16.

Crawford, C. (2004). *Chris Crawford on interactive storytelling*. Pearson Education.

Csikszentmihalyi, M., & Csikzentmihaly, M. (1990). *Flow: The psychology of optimal experience* (Vol. 1990). Harper & Row New York.

Dantec, M., Mantel, M., Lafraire, J., Rouby, C., & Bensafi, M. (2021). On the contribution of the senses to food emotional experience. *Food Quality and Preference*, *92*, 104120.

Decety, J. (1996). Do imagined and executed actions share the same neural substrate? *Cognitive Brain Research*, *3*(2), 87–93.

Decety, J., Jeannerod, M., & Prablanc, C. (1989). The timing of mentally represented actions. *Behavioural Brain Research*, *34*(1–2), 35–42.

Dinh, H. Q., Walker, N., Hodges, L. F., Song, C., & Kobayashi, A. (1999). Evaluating the importance of multi-sensory input on memory and the sense of presence in virtual environments. *Proceedings IEEE Virtual Reality (Cat. No. 99CB36316)*, 222–228.

Draper, J. V, Kaber, D. B., & Usher, J. M. (1998). Telepresence. *Human Factors*, *40*(3), 354–375.

Dużmańska, N., Strojny, P., & Strojny, A. (2018). Can simulator sickness be avoided? A review on temporal aspects of simulator sickness. *Frontiers in Psychology*, *9*, 2132.

Edsall, R., & Larson, K. L. (2006). Decision making in a virtual environment: Effectiveness of a semi-immersive "decision theater" in understanding and assessing human-environment interactions. *Proceedings of AutoCarto*, *6*, 25–28.

Ellis, S. R. (1996). Presence of mind: a reaction to Thomas Sheridan's "further musings on the psychophysics of presence." *Presence: Teleoperators & Virtual Environments*, *5*(2), 247–259.

Erlacher, D., & Schredl, M. (2008). Do REM (lucid) dreamed and executed actions share the same neural substrate? *International Journal of Dream Research*, 7–14.

Ermi, L., & Mäyrä, F. (2005). Fundamental components of the gameplay experience: Analysing immersion. *Proceedings of DiGRA 2005 Conference: Changing Views: Worlds in Play*.

Farley, H. S. (2016). The reality of authentic learning in virtual worlds. *Learning in Virtual Worlds: Research and Applications*, 129–149.

Fauveau, V., Filimonov, A. K., Pyzik, R., Murrough, J., Keefer, L., Liran, O., Spiegel, B., Swirski, F. K., Fayad, Z. A., & Poller, W. C. (2024). Comprehensive assessment of physiological and psychological responses to virtual reality experiences. *Journal of Medical Extended Reality*, *1*(1), 227–241.

Fisher, W. R. (1984). Narration as a human communication paradigm: The case of public moral argument. *Communications Monographs*, *51*(1), 1–22.

Fisher, W. R. (2021). *Human communication as narration: Toward a philosophy of reason, value, and action*. Univ of South Carolina Press.

Fox, J., Ahn, S. J., Janssen, J. H., Yeykelis, L., Segovia, K. Y., & Bailenson, J. N. (2015). Avatars versus agents: a meta-analysis quantifying the effect of agency on social influence. *Human–Computer Interaction*, *30*(5), 401–432.

Fromm, J., Radianti, J., Wehking, C., Stieglitz, S., Majchrzak, T. A., & vom Brocke, J. (2021). More than experience?-On the unique opportunities of virtual reality to afford a holistic experiential learning cycle. *The Internet and Higher Education*, *50*, 100804.

Garau, M. (2003). *The impact of avatar fidelity on social interaction in virtual environments*. University of London, University College London (United Kingdom).

Genette, G. (1980). *Narrative discourse: An essay in method* (Vol. 3). Cornell University Press.

Geng, Y., Wang, X., Nishizawa, W., Negreiro, N., Lin, Y.-J., Atlas, C., Liu-Dujardin, A., Schliemann, F., Gunderson, A., & Crisp, R. (2025). Hyperrealistic VR: A 90-PPD, 1400-Nit, High-Contrast Headset. In *ACM SIGGRAPH 2025 Emerging Technologies* (pp. 1–2).

Gerrig, R. J. (2023). Processes and products of readers' journeys to narrative worlds. *Discourse Processes*, *60*(4–5), 226–243.

Gibson, J. J. (1966). *Gibson , J.J. (1966 ). The senses considered as perceptual systems*. Houghton-Mifflin.

Gilbert, S. B. (2016). Perceived realism of virtual environments depends on authenticity. *Presence: Teleoperators and Virtual Environments*, *24*(4), 322–324.

Godfrey-Smith, P. (1998). *Complexity and the Function of Mind in Nature*. Cambridge University Press.

Goel, L., Junglas, I., Ives, B., & Johnson, N. (2012). Decision-making in-socio and in-situ: Facilitation in virtual worlds. *Decision Support Systems*, *52*(2), 342–352.

Goffman, E. (2002). The presentation of self in everyday life. 1959. *Garden City, NY*, *259*, 2002.

Green, M. C. (2005). *Transportation into narrative worlds: Implications for the self.*

Green, M. C., & Brock, T. C. (2003). In the mind's eye: Transportation-imagery model of narrative persuasion. In *Narrative impact* (pp. 315–341). Psychology Press.

Gunawardena, C. N., & Zittle, F. J. (1997). Social presence as a predictor of satisfaction within a computer-mediated conferencing environment. *American Journal of Distance Education*, *11*(3), 8–26. https://doi.org/10.1080/08923649709526970

Haidt, J. (2003). The moral emotions. *Handbook of Affective Sciences*, *11*(2003), 852–870.

Halliwell, S. (1998). *Aristotle's poetics*. University of Chicago Press.

Hart, S. G. (2006). NASA-task load index (NASA-TLX); 20 years later. *Proceedings of the Human Factors and Ergonomics Society Annual Meeting*, *50*(9), 904–908.

Hart, S. G., & Staveland, L. E. (1988). Development of NASA-TLX (Task Load Index): Results of empirical and theoretical research. In *Advances in psychology* (Vol. 52, pp. 139–183). Elsevier.

Hay, R. (1998). Sense of place in developmental context. *Journal of Environmental Psychology*, *18*(1), 5–29.

Heeter, C. (1992). Being there: The subjective experience of presence. Presence: Teleoperators and Virtual Environments. *Presence: Teleoperators and Virtual Environments, Available at Http://Commtechlab. Msu. Edu/Randd/Research/Beingthere. Html, MIT Press*.

Heeter, C. (2003). Reflections on real presence by a virtual person. *Presence: Teleoperators & Virtual Environments*, *12*(4), 335–345.

Held, R. (1992). Telepresence. *The Journal of the Acoustical Society of America*, *92*(4_Supplement), 2458.

Herd, S., Mingus, B., & O'Reilly, R. (2010). Dopamine and self-directed learning. In *Biologically Inspired Cognitive Architectures 2010* (pp. 58–63). IOS Press.
Herman, D. (2004). *Story logic: Problems and possibilities of narrative*. U of Nebraska Press.
Herman, D. (2009). *Basic elements of narrative*. John Wiley & Sons.
Higgins, J. (1998). Janet H. Murray, Hamlet on the Holodeck: The Future of Narrative in Cyberspace (New York: The Free Press, 1997), 324 pp. ISBN 0-684-82723-9. *Convergence*, *4*(4), 128–130.
Huang, M. P., & Alessi, N. E. (1999). Mental health implications for presence. *CyberPsychology & Behavior*, *2*(1), 15–18.
IJsselsteijn, W. A., De Ridder, H., Freeman, J., & Avons, S. E. (2000). Presence: concept, determinants, and measurement. *Human Vision and Electronic Imaging V*, *3959*, 520–529.
IJsselsteijn, W. A., Freeman, J., Avons, S. E., Davidoff, J., De Ridder, H., & Hamberg, R. (1997). Continuous assessment of presence in stereoscopic displays. *Perception*, *26*(1_suppl), 186.
IJsselsteijn, W., De Ridder, H., Hamberg, R., Bouwhuis, D., & Freeman, J. (1998). Perceived depth and the feeling of presence in 3DTV. *Displays*, *18*(4), 207–214.
(ISPR) International Society for Presence Research. (2000). *The Concept of Presence: Explication Statement*. ISPR. https://ispr.info/about-presence-2/about-presence/
Jabbi, M., Swart, M., & Keysers, C. (2007). Empathy for positive and negative emotions in the gustatory cortex. *Neuroimage*, *34*(4), 1744–1753.
Jeannerod, M. (2001). Neural simulation of action: a unifying mechanism for motor cognition. *Neuroimage*, *14*(1), S103–S109.
Juul, J. (2011). *Half-real: Video games between real rules and fictional worlds*. MIT press.
Kalawsky, R. S., Bee, S. T., & Nee, S. P. (1999). Human factors evaluation techniques to aid understanding of virtual interfaces. *BT Technology Journal*, *17*(1), 128–141.
Kennedy, R. S., Lane, N. E., Berbaum, K. S., & Lilienthal, M. G. (1993). Simulator sickness questionnaire: An enhanced method for quantifying simulator sickness. *The International Journal of Aviation Psychology*, *3*(3), 203–220.
Klimmt, C., & Vorderer, P. (2003). Media psychology "is not yet there": Introducing theories on media entertainment to the presence debate. *Presence*, *12*(4), 346–359.
Kolasinski, E. M. (1995). *Simulator sickness in virtual environments*.
Kolb, D. A. (1984). Experience as the source of learning and development. *Upper Sadle River: Prentice Hall*.
Kyle, G., & Chick, G. (2007). The social construction of a sense of place. *Leisure Sciences*, *29*(3), 209–225.
Lackner, J. R. (2016). Nat Durlach and the Founding of Presence. *Presence*, *25*(2), 161–165.
Lewkowicz, D. J. (2001). The concept of ecological validity: What are its limitations and is it bad to be invalid? *Infancy*, *2*(4), 437–450.
Li, Y., Ch'ng, E., Cobb, S., & See, S. (2021). Presence and Communication in Hybrid Virtual and Augmented Reality Environments. *PRESENCE: Virtual and Augmented Reality*, 1–40.
Lim, S., & Reeves, B. (2010). Computer agents versus avatars: Responses to interactive game characters controlled by a computer or other player. *International Journal of Human-Computer Studies*, *68*(1–2), 57–68.
Lombard, M. (2000). Resources for the study of presence: Presence explication. *Retrieved September*, *3*, 2000.
Lombard, M., & Ditton, T. (1997). At the heart of it all: The concept of presence. *Journal of Computer-Mediated Communication*, *3*(2), JCMC321.

Lombard, M., Ditton, T. B., Crane, D., Davis, B., Gil-Egui, G., Horvath, K., Rossman, J., & Park, S. (2000). Measuring presence: A literature-based approach to the development of a standardized paper-and-pencil instrument. *Third International Workshop on Presence, Delft, the Netherlands*, *240*, 2–4.

Lombardi, M. M., & Oblinger, D. G. (2007). Authentic learning for the 21st century: An overview. *Educause Learning Initiative*, *1*(2007), 1–12.

Majgaard, G., & Weitze, C. (2020). Virtual experiential learning, learning design and interaction in extended reality simulations. *I 14th European Conference on Games Based Learning, ECGBL*, 372–379.

Mania, K., & Chalmers, A. (2001). The effects of levels of immersion on memory and presence in virtual environments: A reality centered approach. *Cyberpsychology & Behavior*, *4*(2), 247–264.

Martens, M. A. G., Antley, A., Freeman, D., Slater, M., Harrison, P. J., & Tunbridge, E. M. (2019). It feels real: physiological responses to a stressful virtual reality environment and its impact on working memory. *Journal of Psychopharmacology*, *33*(10), 1264–1273.

Matheis, R. J., Schultheis, M. T., Tiersky, L. A., DeLuca, J., Millis, S. R., & Rizzo, A. (2007). Is learning and memory different in a virtual environment? *The Clinical Neuropsychologist*, *21*(1), 146–161.

McClean, P. (2001). *Virtual Worlds in Large Enrollment Science Classes Significantly Improve Authentic Learning Phillip McClean, Bernhardt Saini-Eidukat, Donald Schwert, Brian Slator, Alan White North Dakota State University.*

McCloud, S., & Martin, M. (1993). *Understanding comics: The invisible art* (Vol. 106). Kitchen sink press Northampton, MA.

Medley, S., & Haddad, H. (2011). The realism continuum, representation and perception. *The International Journal of the Image*, *1*(2), 145.

Mennecke, B. E., Triplett, J. L., Hassall, L. M., & Conde, Z. J. (2010). Embodied social presence theory. *System Sciences (HICSS), 2010 43rd Hawaii International Conference On*, 1–10.

Milgram, P., & Kishino, F. (1994). A taxonomy of mixed reality visual displays. *IEICE TRANSACTIONS on Information and Systems*, *77*(12), 1321–1329.

Minsky, M. (1980). *Telepresence*.

Montello, D. R., Waller, D., Hegarty, M., & Richardson, A. E. (2004). Spatial memory of real environments, virtual environments, and maps. In *Human spatial memory* (pp. 271–306). Psychology Press.

Murphy, D., & Skarbez, R. (2020). What do we mean when we say "presence"? *PRESENCE: Virtual and Augmented Reality*, *29*, 171–190.

Murray, J. H. (2012). Active creation of belief. *Humanistic Design for an Emerging Medium. Https://Inventingthemedium. Com/2012/01/30/Active-Creation-of-Belief.*

Nabavi, R. T., & Bijandi, M. S. (2012). Bandura's social learning theory & social cognitive learning theory. *Theory of Developmental Psychology*, *1*(1), 1–24.

Nell, V. (1988). The psychology of reading for pleasure: Needs and gratifications. *Reading Research Quarterly*, 6–50.

Okita, S. Y., Bailenson, J., & Schwartz, D. L. (2007). The mere belief of social interaction improves learning. *Proceedings of the Annual Meeting of the Cognitive Science Society*, *29*(29).

Onega, S., & Landa, J. A. G. (2014a). *Narratology: an introduction*. Routledge.

Onega, S., & Landa, J. A. G. (2014b). *Narratology: an introduction*. Routledge.

Panksepp, J. (2004). *Affective neuroscience: The foundations of human and animal emotions*. Oxford university press.

Pekrun, R. (1992). The impact of emotions on learning and achievement: Towards a theory of cognitive/motivational mediators. *Applied Psychology*, *41*(4), 359–376.

Pfister, L., & Ghellal, S. (2018). Exploring the influence of non-diegetic and diegetic elements on the immersion of 2D games. *Proceedings of the 30th Australian Conference on Computer-Human Interaction*, 490–494.

Pimentel, J. R. (1999). Design of net-learning systems based on experiential learning. *Journal of Asynchronous Learning Networks*, *3*(2), 64–90.

Pleban, R. J., Eakin, D. E., Salter, M. S., & Matthews, M. D. (2001). *Training and assessment of decision-making skills in virtual environments*.

Pleban, R. J., Matthews, M. D., Salter, M. S., & Eakin, D. E. (2002). Training and assessing complex decision-making in a virtual environment. *Perceptual and Motor Skills*, *94*(3), 871–882.

Radianti, J., Majchrzak, T. A., Fromm, J., & Wohlgenannt, I. (2020). A systematic review of immersive virtual reality applications for higher education: Design elements, lessons learned, and research agenda. *Computers & Education*, *147*, 103778.

Relph, E. (1976). *Place and placelessness* (Vol. 67). Pion London.

Relph, E. (2007). Spirit of place and sense of place in virtual realities. *Techne: Research in Philosophy & Technology*, *10*(3).

Rettie, R. (2003). *Connectedness, awareness and social presence*.

Ricoeur, P., & Ricoeur, P. (1984). *Time and narrative, Volume 3* (Vol. 3). University of Chicago press.

Riva, G., Davide, F., & IJsselsteijn, W. A. (2003). Being there: The experience of presence in mediated environments. *Being There: Concepts, Effects and Measurement of User Presence in Synthetic Environments*, *5*, 2003.

Rogers, S. (2014). *Level Up! The guide to great video game design*. John Wiley & Sons.

Roth, D., Klelnbeck, C., Feigl, T., Mutschler, C., & Latoschik, M. E. (2018). Beyond replication: Augmenting social behaviors in multi-user virtual realities. *2018 IEEE Conference on Virtual Reality and 3D User Interfaces (VR)*, 215–222.

Ruiz-Martín, H., & Bybee, R. W. (2022). The cognitive principles of learning underlying the 5E Model of Instruction. *International Journal of STEM Education*, *9*(1), 21.

Ryan, M.-L. (2015a). *Narrative as virtual reality 2: Revisiting immersion and interactivity in literature and electronic media*. JHU press.

Ryan, M.-L. (2015b). *Narrative as virtual reality 2: Revisiting immersion and interactivity in literature and electronic media*. JHU press.

Sacau, A., Laarni, J., & Hartmann, T. (2008). Influence of individual factors on presence. *Computers in Human Behavior*, *24*(5), 2255–2273.

Salmond, M. (2021). *Video game level design: how to create video games with emotion, interaction, and engagement*. Bloomsbury Academic.

San Chee, Y. (2001). Virtual reality in education: Rooting learning in experience. *International Symposium on Virtual Education*, *41*.

Schloerb, D. W. (1995). A quantitative measure of telepresence. *Presence: Teleoperators & Virtual Environments*, *4*(1), 64–80.

Schoenau-Fog, H. (2011). The player engagement process–an exploration of continuation desire in digital games. *Proceedings of DiGRA 2011 Conference: Think Design Play*.

Schouten, A. P., van den Hooff, B., & Feldberg, F. (2010). *Real decisions in virtual worlds: Team collaboration and decision making in 3D virtual worlds*.

Schouten, A. P., van den Hooff, B., & Feldberg, F. (2016). Virtual team work: Group decision making in 3D virtual environments. *Communication Research*, *43*(2), 180–210.

Schroeder, R. (2001a). *The social life of avatars: Presence and interaction in shared virtual environments*. Springer Science & Business Media.

Schroeder, R. (2001b). *The social life of avatars: Presence and interaction in shared virtual environments*. Springer Science & Business Media.

Schubert, T., & Crusius, J. (2002). Five theses on the book problem: Presence in books, film and VR. *PRESENCE 2002-Proceedings of the Fifth International Workshop on Presence*, 53–59.

Shamai, S., & Ilatov, Z. (2005). Measuring sense of place: Methodological aspects. *Tijdschrift Voor Economische En Sociale Geografie*, *96*(5), 467–476.

Sheridan, T. B. (1992). Musings on telepresence and virtual presence. *Presence Teleoperators Virtual Environ.*, *1*(1), 120–125.

Skarbez, R., Brooks Frederick P, J., & Whitton, M. C. (2017). A survey of presence and related concepts. *ACM Computing Surveys (CSUR)*, *50*(6), 1–39.

Skarbez, R., Neyret, S., Brooks, F. P., Slater, M., & Whitton, M. C. (2017a). A psychophysical experiment regarding components of the plausibility illusion. *IEEE Transactions on Visualization and Computer Graphics*, *23*(4), 1369–1378.

Skarbez, R., Neyret, S., Brooks, F. P., Slater, M., & Whitton, M. C. (2017b). A psychophysical experiment regarding components of the plausibility illusion. *IEEE Transactions on Visualization and Computer Graphics*, *23*(4), 1369–1378.

Slater, M. (1999). Measuring presence: A response to the Witmer and Singer presence questionnaire. *Presence: Teleoperators and Virtual Environments*, *8*(5), 560–565.

Slater, M. (2003). A note on presence terminology. *Presence Connect*, *3*(3), 1–5.

Slater, M. (2009). Place illusion and plausibility can lead to realistic behaviour in immersive virtual environments. *Philosophical Transactions of the Royal Society B: Biological Sciences*, *364*(1535), 3549–3557.

Slater, M. (2016). Remembering Nat Durlach. *Presence*, *25*(3), 287.

Slater, M. (2018). Immersion and the illusion of presence in virtual reality. *British Journal of Psychology*, *109*(3), 431–433.

Slater, M., & Usoh, M. (1993). Presence in immersive virtual environments. *Proceedings of IEEE Virtual Reality Annual International Symposium*, 90–96.

Slater, M., Usoh, M., & Steed, A. (1994a). Depth of presence in virtual environments. *Presence: Teleoperators & Virtual Environments*, *3*(2), 130–144.

Slater, M., Usoh, M., & Steed, A. (1994b). Depth of presence in virtual environments. *Presence: Teleoperators & Virtual Environments*, *3*(2), 130–144.

Slater, M., & Wilbur, S. (1997a). A framework for immersive virtual environments (FIVE): Speculations on the role of presence in virtual environments. *Presence: Teleoperators & Virtual Environments*, *6*(6), 603–616.

Slater, M., & Wilbur, S. (1997b). A framework for immersive virtual environments (FIVE): Speculations on the role of presence in virtual environments. *Presence: Teleoperators and Virtual Environments*, *6*, 603–616.

Sorden, S. D. (2013). The Cognitive Theory. *The Handbook of Educational Theories*, 155.

Spennemann, R., & Orthia, L. A. (2022). Creating a Market for Technology through Film. *AAA: Arbeiten Aus Anglistik Und Amerikanistik*, *47*(2), 225–242.

Stephenson, N. (1994). *Snow Crash*. Penguin UK.

Steuer, J. (1992). Defining virtual reality: Dimensions determining telepresence. *Journal of Communication*, *42*(4), 73–93.

Stokols, D. (1978). *Environmental psychology*.

Styles, E. (2006). *The psychology of attention*. Psychology Press.

Sweller, J. (2011). Cognitive load theory. In *Psychology of learning and motivation* (Vol. 55, pp. 37–76). Elsevier.

Sweller, J. (2020). Cognitive load theory and educational technology. *Educational Technology Research and Development*, *68*(1), 1–16.

Tajfel, H., Turner, J., Austin, W. G., & Worchel, S. (2001). An integrative theory of intergroup conflict. *Intergroup Relations: Essential Readings*, 94–109.

Tan, E. S. (2013). *Emotion and the structure of narrative film: Film as an emotion machine*. Routledge.

Totten, C. W. (2019). *Architectural approach to level design*. CRC Press.

Truby, J. (2008). *The anatomy of story: 22 steps to becoming a master storyteller*. Farrar, Straus and Giroux.

Turner, P., & Turner, S. (2006). Place, sense of place, and presence. *Presence: Teleoperators & Virtual Environments*, *15*(2), 204–217.

Turner, P., & Turner, S. (2011). The book problem is all in the mind. *International Society for Presence Research Annual Conference, Edinburgh, Scotland*.

Tyng, C. M., Amin, H. U., Saad, M. N. M., & Malik, A. S. (2017). The influences of emotion on learning and memory. *Frontiers in Psychology*, *8*, 235933.

Von der Pütten, A. M., Krämer, N. C., Gratch, J., & Kang, S.-H. (2010). "It doesn't matter what you are!" Explaining social effects of agents and avatars. *Computers in Human Behavior*, *26*(6), 1641–1650.

Vorderer, P. (1993). Audience involvement and program loyalty. *Poetics*, *22*(1–2), 89–98.

Vorderer, P., & Hartmann, T. (2009). Entertainment and enjoyment as media effects. In *Media effects* (pp. 548–566). Routledge.

Waller, D., Hunt, E., & Knapp, D. (1998). The transfer of spatial knowledge in virtual environment training. *Presence*, *7*(2), 129–143.

Walton, K. L. (1980). Appreciating fiction: suspending disbelief or pretending belief? *Dispositio*, *5*(13/14), 1–18.

Wilcox-Netepczuk, D. (2013). Immersion and realism in video games-the confused moniker of video game engrossment. *Proceedings of CGAMES'2013 USA*, 92–95.

Wirth, W., Hartmann, T., Böcking, S., Vorderer, P., Klimmt, C., Schramm, H., Saari, T., Laarni, J., Ravaja, N., & Gouveia, F. R. (2007). A process model of the formation of spatial presence experiences. *Media Psychology*, *9*(3), 493–525.

Wise, R. A. (2004). Dopamine, learning and motivation. *Nature Reviews Neuroscience*, *5*(6), 483–494.

Witmer, B. G., & Singer, M. J. (1998). Measuring presence in virtual environments: A presence questionnaire. *Presence*, *7*(3), 225–240.

Wohlgenannt, I., Fromm, J., Stieglitz, S., Radianti, J., & Majchrzak, T. A. (2019). *Virtual reality in higher education: Preliminary results from a design-science-research project*.

Wong, T. L., Yun, Z., Ambur, G., & Etter, J. (2017). Folded optics with birefringent reflective polarizers. *Digital Optical Technologies 2017*, *10335*, 84–90.

Xiao, L., Zhao, Y., Lindberg, D., Hegland, J., Moczydlowski, S., Penner, E., Tebbs, D., Terpstra, D., Ender, I., & Lin, Y.-J. (2025). Wide Field-of-View Mixed Reality. In *ACM SIGGRAPH 2025 Emerging Technologies* (pp. 1–2).

Yee, N., & Bailenson, J. N. (2007). The Proteus Effect: The Effect of Transformed Self-Representation on Behavior. *Human Communication Research*, *33*, 271–290.